\documentstyle[12pt,epsfig]{article}
\begin{document}
\textwidth 6.75in
\textheight 9in

\begin {center}
{\Large Data on $J/\Psi \to \gamma (K^\pm K^0_S\pi^\mp)$
and $\gamma (\eta \pi ^+\pi ^-)$}

\vskip 3mm
D.V.~Bugg
\footnote{email address: david.bugg@stfc.ac.uk},   \\
{Queen Mary, University of London, London E1\,4NS, UK}
\end {center}
\vskip 2mm

\begin{abstract}
Data on $J/\Psi \to \gamma (\eta \pi ^+\pi ^-)$ and
$\gamma (K^\pm K^0_S\pi ^\mp )$ from 58M $J/\Psi$ hadronic interactions
in the BES II detector are analysed.
They throw new light on $J^P = 0^-$ mesons throughout the mass
range up to 2 GeV, notably $\eta(1440)$.
For the first time, this state is fitted including the full
$s$-dependence of decays to $a_0(980)\pi$, $\eta \sigma$, $K\bar K^*$,
$\kappa \bar K$ and $f_0(980)\eta$, including the dispersive term
associated with this $s$-dependence.
Two types of fit are reported.
The first uses a single $\eta (1440)$.
The second uses separate $\eta (1405)$ and $\eta (1475)$.
The strong $s$-dependence of P-wave production of $K^*K$ shifts the
fitted mass of $\eta (1475)$ down to 1440 MeV.
As a result, the two types of fit give almost identical results and the
conclusion from these data is that there is no significant evidence for
two separate resonances.
In $K\bar K \pi$, there are definite though small contributions from
$f_1(1420)$ and $f_1(1285)$.
The optimum fit requires either a small additional $f_1(1510)$ or
alternatively a small perturbation to the high mass tail of
$f_1(1420)$.
At higher masses, there is a broad $J^P=0^-$ peak in $\eta \pi \pi$ at
1850 MeV.
There is also a weak but definite signal consistent with
$\eta _2 (1870) \to \eta \pi \pi$.

\vspace{5mm}

\end{abstract}

\section {Introduction}
The analysis reported here was done in the years 2002--3 in
collaboration with two senior members of the BES collaboration.
The work was funded by the Royal Society and Queen Mary College
via an official agreement with the Chinese Academy of Sciences
and the BES collaboration.
This agreement guaranteed access to BES 2 data for the purpose of
publications.
Since then, the BES management has refused to appoint internal
referees, hence blocking a collaboration publication and comments from
the collaboration.
The data illuminate the spectroscopy of $J^P=0^-$ mesons
and should be in the public domain.
There is a responsibility to publish work which has been
supported from public funds with approximately $\pounds 60,000$.
The wording of the article is close to a document submitted
late in 2008 to the BES management, except for some
clarification of wording.

There have been many earlier studies
of $\eta (1440)$ in $J/\Psi$ radiative decays, $\bar pp$ annihilation
and $\pi N$ interactions.
Extensive references are to be found in the
listings of the Particle Data Group (PDG) \cite {PDG}.
Their review concludes that two nearby $0^{-+}$ resonances are required
in the mass range 1400--1500 MeV.
The lower one is at 1405 MeV, decaying to $\eta \pi \pi$ and
$\kappa \bar K$, where $\kappa$ stands for the $K\pi$ S-wave.
The upper state is at 1475 MeV, decaying dominantly to $K^*(890)\bar K$.

However, almost all of these analyses are based on fitting the strong
$K^*(890)\bar K$ channel with a Breit-Wigner resonance of constant
width.
The threshold for this channel is at 1390 MeV, although smeared
out to some extent by the 50 MeV full-width of the $K^*$.
The $K^*\bar K$ channel with $J^P = 0^-$ is produced in a P wave.
As a result, the intensity of the $K^*\bar K$ channel increases
with kaon momentum $k$ as $k^3$.
This rapid $s$-dependence enhances  the upper side of the $K^*\bar K$
peak strongly.
It also gives rise to a strong dispersive correction in the Breit-Wigner
denominator, as described in subsection 4.3.
Only one previous analysis includes the first feature and none contains
the second.
They are therefore prejudiced towards two separate resonances.

As well as the possible $\eta (1405)$ and $\eta (1475)$, there is
evidence from four experiments on $\pi p \to (\eta \pi \pi )p$ for
$\eta (1295)$ \cite {Fukui}, \cite {Alde}, \cite {Manak} and \cite
{Adams}.
In these data, the possibile exchange of both natural and
unnatural parity mesons requires a rank 2 analysis, with partial
interference between the two alternatives, hence incomplete
interferences between $f_1(1285)$ and $\eta (1295)$.
The possibility of no interference between $f_1(1285)$ and $\eta (1295)$
leaves some room for uncertainty in conclusions.
There is no definite evidence for $\eta (1295)$ in $J/\Psi $ decays
and only weak evidence in $\bar pp$ annihilation \cite {Nana}.

There are therefore three controversial candidates for two
radial excitations of $\eta $ and $\eta '$.
Some people have argued that $\eta (1295)$ has been confused
with $f_1(1285)$ \cite {KlemptZ}.
Others argue from the strong production of $\eta (1405)$ and
$\eta (1475)$ in $J/\Psi$ radiative decays that one of them is the
$0^-$ glueball \cite {Masoni}.
However, Lattice Gauge calculations predict the glueball above
2 GeV \cite {Morningstar}.

One must remember too that the $\eta $ and $\eta '$ are
well established to have compositions
\begin {eqnarray}
\eta &=& n\bar n
 \cos \phi - s\bar s \sin \phi \\
\eta '&=& n\bar n \sin \phi + s\bar s \cos \phi,
\end {eqnarray}
where $n\bar n = (u\bar u + d\bar d)/\sqrt {2}$ and $\phi =
41.5^\circ$  if there is no glueball component \cite {Escribano}.
In  $J/\Psi$ radiative decays, the $\eta '$ is produced with an
intensity a factor 5 larger than the $\eta$.
Similar mixing for radial excitations could explain surprisingly strong
production of $\eta (1405)$ and/or $\eta (1475)$ in $J/\Psi$ decays and
$\bar pp$ annihilation.

There are two likely scenarios.
The first is that the $\eta (1295)$ does not exist and $\eta (1405)$
may be assigned dominantly as $\bar nn$ and $\eta (1475)$ as $\bar ss$.
The second scenario is that $\eta (1295)$ does exist and $\eta
(1405)$ and $\eta (1475)$ are really one resonance.
These two alternatives are fitted to data presented here.
The $s$-dependence of the Breit-Wigner amplitudes is treated fully
for the first time.

The higher mass range for $\eta \pi \pi$ is also of considerable
interest.
BES II data provide evidence for a 7 standard deviation signal
in $\eta '\pi \pi$ for a narrow $X(1835)$ with $\Gamma = 68$ MeV
\cite {X1835}.
It is important to know if it appears in
$\eta \pi \pi$ and if so with what branching fraction.

Section 2 presents technical details of the selection of data for the
channels $\gamma (\eta \pi ^+\pi ^-)$, $\gamma (\eta K^+K^-)$ and
$\gamma (K^\pm K^0_S\pi ^\mp)$ and discusses backgrounds.
The $\gamma (\eta K^+K^-)$ channel has not been examined before.
It is useful in establishing the magnitude of possible
$f_0(980)$ contributions to $\eta K \bar K$.
Sections 3 and 4 present features of the data and the formulae needed to
treat the $s$-dependence of Breit-Wigner amplitudes.
Section 5 gives details of the fit to the mass range up to 1600 MeV.
This is the centre-piece of the paper.
Section 6 considers the high mass range above 1600 MeV and estimates
the magnitude possible for $X(1835) \to \eta \pi \pi$;
the low value observed is inconsistent with the $X(1835)$ being
dominantly $\bar nn$.
Section 7 returns to the interpretation of $\eta (1440)$ and
$f_1(1510)$.
Section 8 summarises conclusions.

\section {Data selection}
The data fitted here were taken with the BES\,II detector.
Full details of the detector and its upgrade are reported by
Bai et al. \cite {DetectA}, \cite {DetectB}.
It has cylindrical symmetry around the intersecting $e^+e^-$ beams.
Its essential features are
(i) a Main Drift Chamber (MDC) for the measurement of charged particles,
(ii) time of flight detectors with a $\sigma$ of 180 ps, and
(iii) a 12 radiation length Barrel Shower Counter comprised of
gas proportional tubes interleaved with lead sheets.
The MDC measures $dE/dx$.
Together with the time of flight detectors, it separates $\pi$ and $K$
up to $\sim 700$ MeV/c.
Outside the $\gamma$ detectors is a magnet providing a field of
0.4T.
This magnet is instrumented with muon detectors, but for  present
work they serve simply to reject $\mu ^+\mu ^-$ pairs.
The Main Drift Chamber provides full coverage of charged particles for
lab angles $\theta$ with $|\cos \theta | < 0.84$.

\subsection {Selection of $\gamma (\eta \pi ^+\pi ^-)$
and $\gamma (\eta K^+K^-)$}
These events have been selected by demanding exactly three photons and
two charged tracks with balancing charges.
Both charged particles must be positively identified by time-of-flight
and/or $dE/dx$ as pions (for $\eta \pi ^+\pi ^-$) or as kaons
(for $\eta K^+K^-$).
The vertex must lie within a cylinder of 1.5 cm radius, 30 cm long,
centred on the intersecting $e^+e^-$ beams.
Two photons must have an invariant mass within 70 MeV of the $\eta$.
Events are rejected if any two photons
produce a mass within 40 MeV of the $\pi ^0$.
Photons may also originate from interaction of charged tracks in the
detector.
Any photons lying within a $6^\circ$ cone around a charged track are
discarded.

Candidate events have been subjected to 4-constraint kinematic fits to
$3\gamma \pi ^+\pi ^-$ and $5C$ fits to $\gamma \eta \pi ^+\pi ^-$
(or correspondingly $3\gamma K^+K^-$ and $\gamma \eta K^+K^-$); they
are accepted only if the $\chi ^2$ probability of each fit is $>5\%$.
Any events where two photons are consistent with the final state
$2\gamma \eta \pi ^+\pi ^-$ or $2\gamma \eta K^+K^-$ are rejected if
$\chi ^2 <20$.
They are also rejected if the measured particles are consistent
with the loss of one photon and a fit to $4\gamma \pi ^+\pi ^-$ or
$4\gamma K^+K^-$ with $\chi ^2 <20$.

\begin {figure}  [htb]
\begin {center}
\vskip -14mm
\epsfig{file=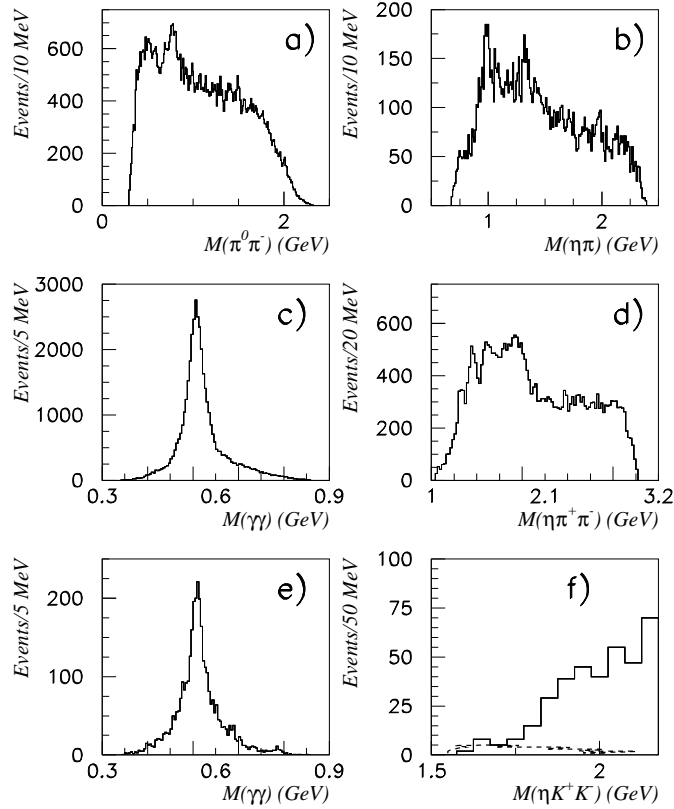,width=9.25cm}\
\vskip -6mm
\caption{Mass distributions for (a)  $\pi ^0\pi ^\pm$ and (b)
$\eta \pi ^\mp$ for events fitted to $\pi ^0\eta \pi ^+\pi ^-$
after selecting events with $M(\pi ^0\pi ^\pm)$ in the range 0.67 to
0.87 GeV; (c) mass distributions for the $\gamma \gamma$ pair closest to
the $\eta$ for $3\gamma \pi ^+\pi ^-$ and (e) $3\gamma K^+K^-$;
(d) $M(\eta \pi ^+\pi ^-)$ after a kinematic fit to $\gamma \eta \pi ^+
\pi ^-$; (f) $M(\eta K^+K^-)$ after a kinematic fit to
$\gamma \eta K^+K^-$; the dashed curve shows the predicted mass
distribution for $f_0(980) \to K^+K^-$ from $\eta (1440)$.}
\end {center}
\end {figure}

The main background arises from the loss of one photon from the final
state $\eta \pi ^+\pi ^-\pi ^0$, which is dominated by
$a_2(1320)\rho$.
This background has been studied using the standard SIMBES Monte Carlo
simulation of the detector.
After the loss of one soft photon, it leads to a strong peak at
high $\eta \pi \pi$ masses 2300-2800 MeV.
This background may be reconstructed from data by assuming one
missing photon in the kinematic fit.
For events fitted this way, Fig. 1(a) displays the $\pi ^0\pi ^\pm$
mass distribution.
There is a clear peak due to $\rho (770)$.
If $M(\pi ^0\pi  ^\pm)$ lies in the mass range 670--870 MeV, one
finds the $\eta \pi ^\mp$ mass distribution of Fig. 1(b).
It shows clear $a_0(980)$ and $a_2(1320)$ peaks.
In order to suppress these backgrounds further, the missing momentum is
calculated from the measured $\eta \pi ^+\pi ^-$ of each
candidate event.
Assuming the missing particle to be a $\pi ^0$, any event with
$M(\pi ^0\pi ^\pm)$ in the mass range 670-870 MeV is rejected.
However, a minor problem is that events lying in the charge zero
peak of the $\rho$ cannot be eliminated without rejecting required
$\gamma \eta \sigma$ events, where $\sigma$ is a shortand for the $\pi
\pi$ S-wave.

Fig. 1(c) shows the mass distribution for the $\gamma \gamma$ pair
closest to the $\eta$ for all combinations fitting
$3\gamma \pi ^+\pi ^-$.
Fig. 1(e) shows the corresponding plot for candidate $\eta K^+K^-$
events.
Fig. 1(d) shows the $\eta \pi \pi$ mass distribution from the
final data sample.
Peaks are visible at 1285 and 1415 MeV, followed by
a sharp rise at 1490 MeV to a shoulder at $\sim 1550$ MeV.
There is a further peak at  $\sim 1850$ MeV, then a rapid fall to 2
GeV.

Above 2 GeV, there is a large contamination from the
background explained above.
The Monte Carlo simulation of the $a_2\rho$  background is fitted to
events in the mass region 2300-2800 MeV, hence
establishing its magnitude.
It agrees with the PDG branching fraction of $J/\Psi \to a_2(1320)\rho$
within its quoted error of $35\%$.

The physics analysis will be restricted to the mass range below 2 GeV.
Since $a_2(1320)\rho$ is the dominant background by a
factor 10, the background at low $\eta \pi ^+\pi ^-$
masses is taken from the Monte Carlo simulation of this channel.
Integrated up to an $\eta \pi \pi$ mass of 2 GeV, this
background is 18.2\%.
Its distribution with $\eta \pi \pi$ mass is illustrated below in
Fig. 7(d).
In the mass range below 1600 MeV it is small and featureless.
Although there is a background of misidentified events,
direct feedthrough from $a_2\rho$ produces at
most 33 events in $\rho \to \pi ^+\pi ^-$, mostly above 1600 MeV.
This small $a_2(1320)$ background surviving in data will be pointed out
later.

At the end of the data selection, there are 18,751 events up to an
$\eta \pi \pi$ mass of 2000 MeV of which an estimated 3350 are
background.
The average efficiency for data reconstruction is $13.5\%$.
The branching fraction is
\begin {equation}
BR[J/\Psi \to \gamma (\eta \pi \pi )] = (7.5 \pm 0.8(syst)) \times
10^{-3}
\end {equation}
up to a mass of 2 GeV, after correction for all charge combinations and
background.
The error is mostly systematic.
It covers the estimated background and the uncertainty in
partial wave amplitudes due to loss of events through the $35^\circ$
holes in the detector around the beam entrance and exit.
The PDG quotes an average branching fraction
$(6.1 \pm 1.0) \times 10^{-3}$.
However, this refers to a Crystal Ball measurement \cite {Edwards}
claiming a peak in $\eta \pi \pi$ at 1700 MeV, substantially
different to what is observed here.

\subsection {Data on $J/\Psi \to \gamma (\eta K^+K^- )$}
For this final state, the main background arises from
$\pi ^0\eta K^+K^-$ after the loss of one photon.
The contamination is estimated to be $\sim 27\%$ from a study of
side-bins either side of the $\eta$ peak, together with
a Monte Carlo estimate of the shape of the background.
The reconstruction efficiency is 7.6\%
and varies little with  $\eta K\bar K$ mass.
Fig. 1(f) shows the mass distribution of events fitting as
$\gamma \eta K^+ K^-$.
The number of events is too small to allow a meaningful physics
analysis.
However, the data are useful in establishing an upper limit on the
number of events due to the final state $f_0(980)\eta$,
$f_0 \to K^+K^-$.
The essential point is that the number of signal events is a factor 100
smaller than for $\gamma \eta \pi ^+\pi ^-$.
After background subtraction, there are 22 events up to an
$\eta K\bar K$ mass of 1.8 GeV and 125 up to 2 GeV.
The latter corresponds to a branching fraction
\begin {equation}
BR[J/\Psi \to \gamma \eta K\bar K] = (1.4 \pm 0.4) \times 10^{-4}
\end {equation}
up to 2 GeV.
Errors are again systematic, from the acceptance.

\subsection {Selection of $\gamma (K^\pm K^0_S\pi ^\mp)$ events}
These events have been selected initially with at least one photon and
with 4 charged tracks having balancing charges.
One and only one charged kaon must be identified by time-of-flight or
$dE/dx$; at least two charged tracks must be positively identified as
pions.

Events are then subjected to a $4C$ kinematic fit, requiring a
$\chi^2$ probability $>5\%$.
The mass of one $\pi ^+\pi ^-$ pair must lie within 25 MeV of the
$K^0_S$.
This pair must be consistent with a vertex within 6 cm of the
second vertex defined by the other two charged particles.
Fig. 2 shows the mass distribution for the reconstructed $K^0_S$
before the kinematic fit.
There is a clean peak with only
a $4\%$ background, which can be evaluated accurately.
Events are rejected if the spectator pion and either of the other two
particles (fitted as pions) have an invariant mass within 25 MeV of
the $K^0_S$.

\begin {figure}  [htb]
\begin {center}
\vskip -16mm
\epsfig{file=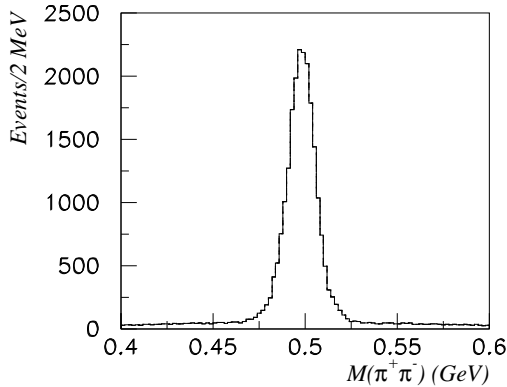,width=7cm}\
\vskip -6mm
\caption{The mass distribution for the reconstructed $K^0_S$.}
\end {center}
\end {figure}

Events are rejected if there is a better fit to final states
$K^{\pm}K^0_S\pi ^{\mp}$ or $K^{\pm}K^0_S\pi ^{\mp}\pi ^0$
without an additional photon.
In order to reject the second process more strongly, events
are discarded if the missing mass attributed to neutral particles
is within 50 MeV of the $\pi ^0$ mass.

The kinematic fit is made assuming that the $K^0_S$ decays at the
primary vertex; the effect of this assumption on the selection of
events is simulated by a Monte Carlo study.
There is a small loss of events, but no significant bias to
kinematics.
Alternative selections have been made fitting two separate
vertices.
These lead to a larger loss of events because of the
difficulties of reconciling two vertices with the
kinematics; within statistics, there is no change to
physics conclusions.

At the end of the data selection, there are 5638 events including an
estimated background of 1068.
The background is substantially smaller over the reduced mass range up
to 1600 MeV.
The average efficiency for data  reconstruction is $10.8\%$.
The resulting branching ratio, corrected for  charge combinations, is
\begin {equation}
BR[J/\Psi \to \gamma K\bar K \pi ] = [2.2  \pm 0.3(syst)]\times 10^{-3},
\end {equation}
up to a mass of 1800 MeV.
The error is almost entirely systematic.
A $10.3\%$ error arises from extrapolating the partial wave analysis
over the holes of the detector, $7.5\%$ uncertainty in the effects of
cuts, and $4.5\%$ in the background; these errors are added in
quadrature.
The PDG quotes an average of $(2.8 \pm 0.6) \times
10^{-3}$, though this is derived from four values spanning the range
$(1.66-4.3)\times 10^{-3}$.

\begin {figure}  [htb]
\begin {center}
\vskip -16mm
\epsfig{file=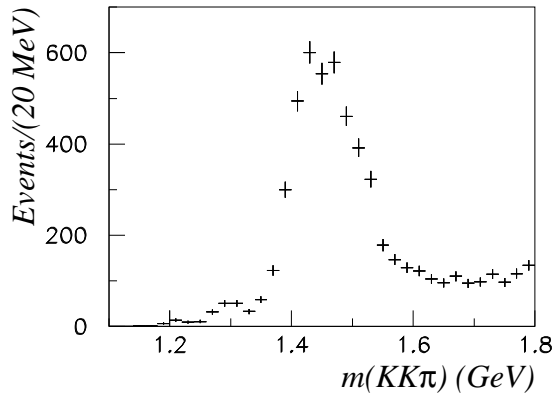,width=8.5cm}\
\vskip -6mm
\caption{The $K^\pm K^0_SK^\mp$ mass spectrum.}
\end {center}
\end {figure}

The Monte Carlo used here to simulate the BES detector was the
standard SIMBES package released in February 2002, in its version
as of March 2003.
During discussions with the BES management, they have claimed that
alterations to the calibration of the time-of-flight system after
March 2003 could affect the selection of events.
This point is discussed in detail in the accompanying paper on
$\gamma (\pi ^+\pi ^-\pi^+\pi ^-)$ data \cite {g4pi}.
In that analysis, no systematic discrepancy is observed between
$(dE/dx)$ and time-of-flight, and the same is true for present data;
no contamination of $\gamma \pi ^+\pi ^-\pi ^+\pi ^0$ events by
$\gamma K^+K^-\pi ^+\pi ^-$ is observed above the $1\%$ level.
In the present data, there is again no observed confusion between
$\pi ^+\pi ^-$ and $K^+K^-$.
For $\gamma \eta \pi ^+\pi ^-$ data, contamination from
$\gamma \eta K^+K^-$ is obviously $<1\%$ because of the small observed
branching fraction for that channel.
There is no evidence for contamination from $\pi ^0\eta K^+K^-$, which
is kinematically well separated from $\gamma \eta \pi^+\pi^-$;
such contamination is estimated to be $<1\%$.
For $\gamma (K^\pm K^0_S\pi^\mp)$ data, the selection of the
$K^0_S$ limits possible background to $4\%$.
The observed background under the $K^0_S$ is consistent with
that originating from $\pi ^0K^\pm K^0_S\pi^\mp$; it rises with
$K\bar K\pi$ mass as expected for that source.

\section {Features of the data}
Fig. 3 shows the $K\bar K\pi$ mass distribution in $J/\Psi \to
\gamma (K^\pm K^0_S\pi ^\mp )$.
A small $f_1$ peak is visible at 1285 MeV.
Then there is a strong but asymmetric peak at $\sim 1450$ MeV on a
slowly rising background.
There may be a small structure causing a drop at
1530 MeV, though this could be a large statistical fluctuation.
A similar shoulder is visible at 1510 MeV in
Mark III data of Bai et al. \cite {BaiM3}.
Several alternative prescriptions for fitting this shoulder will be
described.

The raw mass distribution for $\eta \pi ^+\pi ^-$ is shown in Fig.
1(d) and fits to it will be shown in detail below in Figs. 7 and 11.
There is a narrow peak at $1415 \pm 10$ MeV with a full width of 60 MeV.
There is also a sharp rise at 1500 MeV.
It is vital to realise from the outset that these structures
are superimposed on a large and broad $0^-$ signal extending over the
whole mass range from 1200 to 2000 MeV and interfering coherently,
see Fig. 7(b) below.
Without the inclusion of this broad component, fits fail completely
to describe the $\eta \pi \pi$ mass projection.
In particular, the broad component plays a major role in fitting the
shoulder at 1550 MeV.

The $\gamma 4\pi$ data contain a similar broad peak at 1600 MeV
in $\rho \rho$ with $J^P = 0^-$ \cite {g4pi}.
It is natural that some $\rho \rho$ pairs will de-excite to $\eta
\sigma$ or $a_0(980)\pi$ with $L = 1$ in the decay.
It is also possible that the $\rho \rho$ peak at 1600 MeV is resonant.

\begin {figure}  [htb]
\begin {center}
\vskip -8mm
\epsfig{file=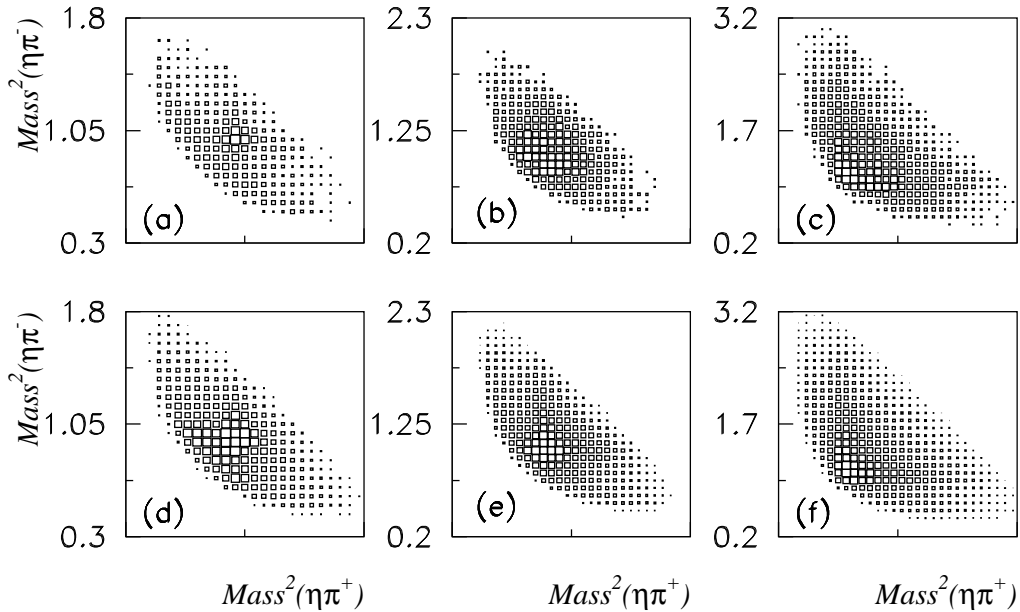,width=14cm}\
\vskip -6mm
\caption{Dalitz plots for $\eta \pi \pi$: (a)--(c) show
data and (d)--(f) the fit. Mass ranges for $\eta \pi \pi$
are 1350--1480 MeV in (a) and (d), 1480--1620 MeV in (b)
and (e), 1620--1950 MeV in (c) and (f).}
\end {center}
\end {figure}

Dalitz plots for $\eta \pi ^+ \pi ^-$ are shown for three mass
ranges in Fig. 4.
The upper 3 panels are for data and the second three from the
fit to $\eta (1440)$ reported below.
The lower set does not contain the statistical fluctuations of
the upper set and the plotting procedure makes it appear that there
are some differences between data and fit.
However, a check on all bins shows no evidence for systematic
discrepancies.
The main feature in all three cases is a peak due to two crossing
$a_0(980)$ bands.

\begin {figure}  [htb]
\begin {center}
\vskip -6mm
\epsfig{file=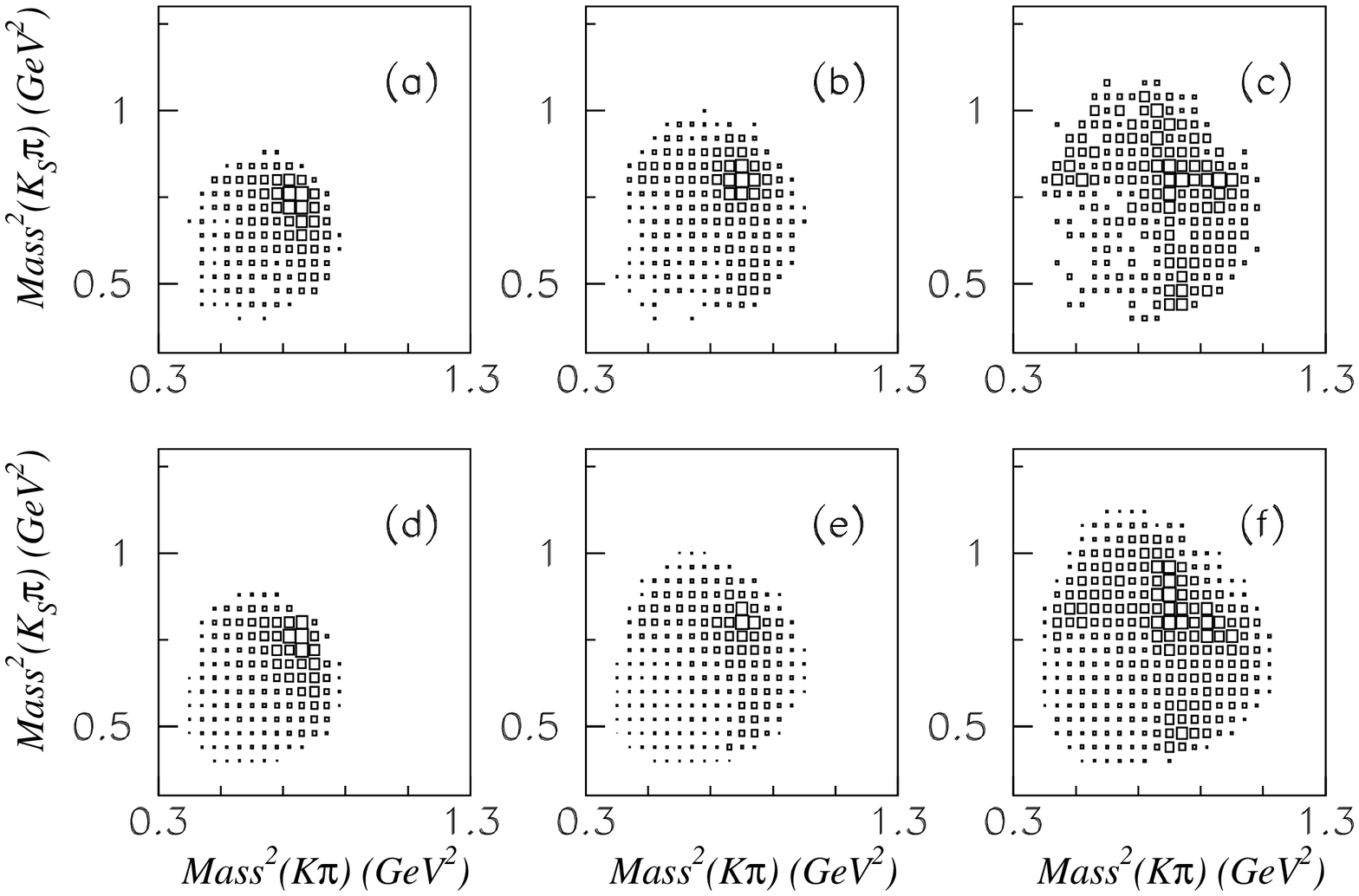,width=14cm}\
\vskip -6mm
\caption{Dalitz plots for $K^\pm K^0_S\pi ^\mp$: (a)--(c) show
data and (d)--(f) the fit.
Mass ranges for $K\bar K\pi$ are 1380--1440 MeV in (a) and (d),
1440--1500 MeV in (b) and (e), 1500--1560 MeV in (c) and (f).}
\end {center}
\end {figure}

Dalitz plots for $K\bar K\pi$ are shown in Fig 5 for three mass
ranges.
Here, the main feature is due to two crossing $K^*(890)$ bands.
There is a strong enhancement near the intersection of these
bands in Figs. 5(a) and (b).
It is much reduced in Fig. 5(c).
This is due to the zero in the $K\pi$ amplitude near the middle of
the $K^*$ band, arising from the angular dependence of the partial
wave amplitude in the decay.

\section {Partial wave analysis}
\subsection {Preliminaries}
The aim of the analysis is to fit the data in terms of conventional
resonances, plus the broad interfering $J^P = 0^-$ component
observed both here and in the $\gamma (\pi ^+\pi ^-\pi ^+\pi ^-)$ data.
It would be good if the spin composition of $\eta \pi \pi$ and
$K\bar K\pi$ could be determined in narrow slices of mass, and this has
been tried.
The main conclusion from that attempt is that $0^-$ is strongly
dominant for both sets of data.
In $K\bar K\pi$ data, contributions from $J^P = 1^+$ are
identified clearly in the narrow mass ranges of $f_1(1285)$ and
$f_1(1420)$.
There is also definite evidence for a small $1^+$
contribution continuing above $f_1(1420)$, up to $\sim 1550$ MeV.
However, it is sufficiently small that its line-shape cannot be
determined with any precision from  slice fits.
The contribution from $J^P = 2^-$ is small.

There are however two difficulties  with trying to analyse the data in
slices.
The first is that the only available information is a differential
cross-section.
This does not allow independent determinations of amplitude and phase.
A related difficulty is that there are complex interferences between
$a_0(980)\pi$, $\eta \sigma$ and $f_0(980)\eta$ in $\eta \pi \pi$
data and likewise in $K\bar K\pi$ data between $K^*\bar K$,
$\kappa \bar K$ and $a_0(980)\pi$, $a_0(980) \to K\bar K$.
In the absence of any strong interferometer in the data with known
quantum numbers, the slice analysis cannot sort out these
interferences.
This leaves no choice but to relate magnitudes and phases through
Breit-Wigner amplitudes fitted directly to the data.

Partial waves which are tried are $J^P = 0^-$,
(orbital angular momentum $L' = 1$ in the production process
$J/\Psi \to \gamma + X$), $1^+$ ($L' = 0$), $2^-$ ($L' = 1$) and
$2^+$ $(L' = 0$).
Of these, $J^P = 2^+$ is negligible.
Decay channels which are considered are
$a_0(980)\pi$, $K^*(890)\bar K$, $\kappa \bar K$, $a_2(1320)\pi$,
$f_2(1270)\eta$, $f_0(980)\eta$ and $\eta \sigma$.

There is an immediate question how $\sigma$ and $\kappa$ should be
parametrised.
In $\pi \pi$ and $K\pi$ elastic scattering, there are Adler zeros
which suppress the amplitudes strongly near threshold.
This is well understood in terms of Chiral Symmetry Breaking, which
creates the Adler zeros.
However, in many sets of data, it is established that $\sigma$ and
$\kappa$ appear as low mass peaks.
The origin of this effect is straightforward.
If the elastic amplitude is written as $N(s)/D(s)$, the denominator
must be universal.
However, the numerator can be quite different in elastic scattering
and production reactions.
BES II data for $J/\Psi \to \omega \pi ^+\pi ^-$ \cite {wpp} may be
fitted accurately with $N(s)$ a constant.
This is also true for BES data on $J/\Psi \to K\bar K\pi \pi$,
where there is again a low mass peak in the $K\pi$ S-wave due to the
$\kappa$ pole \cite {KappBES}.
The present work uses the  $\kappa$ amplitude fitted in Ref. \cite
{JKappa} to LASS data for elastic scattering \cite {Aston}, BES II,
and E791 data.

A problem is that it is not yet fully understood why or how the Adler
zero disappears from the numerator of the amplitude in some production
processes.
In principle, there could be contributions with a numerator
$N(s) = A + Bs$ where $A$ and $B$ are complex constants.
However, this unfortunately leads to large destructive
interferences between the two terms.
This is a well known symptom of over-fitting the data, so
final fits must choose between $\sigma$ and $\kappa$ poles
or alternatively amplitudes the same as for elastic scattering.
Empirically, the present $\gamma (K\bar K\pi )$ data give a considerably
better fit (by 46 in log likelihood) with the $\kappa$ pole amplitude,
i.e. a constant numerator $N(s)$.
[Log likelihood is defined here so that it changes by 0.5 for a one
standard deviation change in the fit].
The $\gamma (\eta \pi ^+\pi ^-)$ data show less discrimination, but
still prefer a fit with the $\sigma$ pole, i.e. no Adler zero in
$N(s)$.
This fit is better than one using the $\pi \pi$ elastic scattering
amplitude by 16 in log likelihood.
Fits to present data are almost identical using the formula for the
$\sigma$ of Zou and Bugg \cite {ZouB} or a recent formula
fitted to the $\sigma$ pole \cite {sigpole}.
The first formula is simpler and is used here.
Both $\sigma$ and $\kappa$ are fitted using a constant for $N(s)$.

The fit to data  is made using 200K Monte Carlo events for $\gamma
(\eta \pi ^+\pi ^-)$ and $52$K for $\gamma (K^\pm K^0_S\pi ^\mp )$ to
simulate detector acceptance.
These Monte Carlo events are generated with the standard SIMBES
package.
All figures show experimental results uncorrected for acceptance.
These corrections are included automatically into the log
likelihood fit.
The acceptance varies only slowly across figures.

A technical detail is that partial wave amplitudes are written in
terms of relativisic tensors; explicit formulae are given in \cite
{formulae}.
In $J/\Psi $ radiative decays, an important point is that
the dominant production process is expected to be via intermediate
$\gamma c \bar c$.
Because of the high mass of these quarks and the interaction of the
photon at a point, the production process is expected to be almost
pointlike.
Data on $J/\Psi \to \gamma 4\pi$ in the accompanying paper verify
this \cite {g4pi}.
They contain a dominant $J^P = 0^-$ signal in $\rho \rho$
with a dependence on $4\pi$ mass  close to the expected $P^3_\gamma$
for $0^-$ production with $L' = 1$; here $P_\gamma$ is the photon lab
momentum.
The consequence for present data is that production should proceed
only through the lowest available orbital angular momentum $L'$.
There are well identified $f_1(1285)$ and $f_1(1420)$ signals in
$\gamma (K\bar K\pi )$ final states.
In both cases, any amplitudes with $L' = 2$ are negligible,
producing improvements in log likelihood $<4$.
For production of $0^-$ states, there is likewise no evidence
for $L' = 3$.
With this simplification, there are distinctive differences between
final states with $J^P = 0^-$, $1^+$, $2^-$ and $2^+$.

\subsection {Details of partial wave fits}
Fits to the 1450 MeV $K\bar K\pi$ peak immediately
reveal the need for both $J^P = 1^+$ and $0^-$ amplitudes.
The $f_1(1420)$ amplitude is conspicuous from its $\cos \theta _d$
dependence, where $\theta _d$ is the decay angle in the rest frame of
the $K^*$ to $K$ and $\pi$.
A free fit gives $M = 1429 \pm 3$ MeV,
compared with the value $1426.4 \pm 0.9$ MeV of the PDG.
If the $f_1(1420)$ is removed from the final fit, log likelihood is
worse by $\ge 129$ for two less fitting parameters.
Statistically  this is nearly a 14 standard deviation effect.

There is also a small but definite $f_1(1285)$ contribution,
decaying to $\kappa \bar K$.
If the $\eta (1295)$ is included in the fit, its contribution is
small and will be discussed in Section 7.

Earlier analyses have freqently fitted the $a_0(980)$ freely to
$K\bar K \pi$ data.
The $a_0(980)$ contribution lies along the upper right-hand edge of
Dalitz plots of Fig. 5.
For low $K\bar K\pi$ masses, it interferes strongly with the two
$K^*(890)\bar K$ components, see Figs. 5(a) and (b).
It can easily be confused with the intersection between the two
$K^*(890)$ bands.
The $a_0(980)\pi$ contribution in $\eta \pi \pi$ data does not
suffer from this problem.
Fits have therefore been made constraining the decay branching ratio of
$a_0(980)$ between $\eta \pi$ and $K\bar K$ to a value consistent with
the Flatt\' e form for $a_0(980)$ \cite {a0980}.
Integrated over the $a_0(980)$ peak up to 1.3 GeV, the branching
fraction to $K\bar K$ is $(24 \pm 2)\%$ of that in $\eta \pi$.
This constraint results in small but well defined contributions of
$a_0(980)$ to $K\bar K\pi$ data.
Production phases are fitted freely.

\subsection {Parametrisation of amplitudes}
It is essential to include into the Breit-Wigner amplitude
the strong $s$-dependence of the phase space for P-wave
production of $K^*\bar K$.
This phase spaces rises as $k^3$, where $k$ is the kaon momentum
in the $K^*K$ rest frame.
As a reminder, the intensity for a two-channel resonance is
\begin {equation}
I(s) = \frac {M^2\Gamma _2(s)}
{(M^2 - s)^2 + M^2\Gamma ^2_{tot}(s)},
\end {equation}
where $\Gamma _{tot}$ is the total width.
The factor $\Gamma _2(s)$ in the numerator arises from the phase space
of the final $K^* K$ state.
The same $s$-dependence {\it must} be included into the denominator,
otherwise an inconsistency with unitarity arises between numerator
and denominator.
The correct form for the Breit-Wigner amplitude for
decay to channel $j$ (for example $K^*\bar K$) is
\begin {equation}
f_j = \frac {F_jBj\rho _j(s)}
{M^2 - s - m(s) - i\sum _ig^2_i\rho _i(s)F^2_iB^2_i},
\end {equation}
where $g_i$ is the coupling constant for each channel
$i$ and $\rho (s)$ is its phase space.
A form factor $F$ is included, and (where required) a centrifugal
barrier factor $B$.
Details will be given shortly.
The evaluation of phase space is done numerically, using Eq.
(40) of Ref. \cite {BSZ} which will be repeated here for
completeness.
As an example,
the 3-body phase space for $K^*\bar K$ is given by
\begin {equation}
\rho _{K^*\bar K}(s) = \int ^{(\sqrt {s} - m_K)^2} _{4m^2_K}
\frac {ds_1}{\pi} \frac {4|k||k_1|}{\sqrt {s s_1}}
|T_{K^*\bar K} (s_1)|^2 ,
\end {equation}
where $T$ is the Breit-Wigner amplitude for the $K^*$.
Here $s$ refers to the resonance $X$ and $k$ to the momentum of the
$K^*$ or $K$ in the rest frame of $X$; $s_1$ and $k_1$ refer to
the $K^*$ and the momenta of decay $K$ and $\pi$ in its rest frame.
For cases where there is angular momentum in the decay,
the centrifugal barrier factor $ B$ appears in the numerator.

\begin {figure}  [htb]
\begin {center}
\vskip -6mm
\epsfig{file=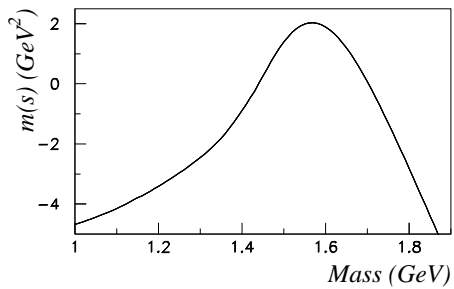,width=7cm}\
\vskip -6mm
\caption{The dispersive term $m(s)$ for $\eta (1440)$.}
\end {center}
\end {figure}

The term $m(s)$ is a dispersive correction to the real part of the
amplitude, required to make the amplitude fully analytic:
\begin {equation}
m(s) = \frac {M^2 - s}{\pi }\int \frac {M\Gamma _{tot}(s') ds'}
{(s' - s)(M^2 - s')}.
\end {equation}
It includes a subtraction on resonance making the integral
strongly convergent.
The resulting $m(s) $ for $\eta (1440)$ is illustrated in Fig. 6.
In the 1440 MeV mass region, the term $m(s)$ actually varies more
strongly with $s$ than the term $(M^2 - s)$ in the Breit-Wigner
denominator, leading to major deviations of phases from those of
a Breit-Wigner resonance of constant width.
This is one of the essential new features of the present analysis.

In fitting $\eta \pi \pi$ data, it is unavoidable to introduce a broad
$J^P=0^-$ signal.
It needs to accomodate peaks observed in $J/\Psi \to \rho \rho$,
$\omega \omega$, $K^*\bar K^*$ and $\phi \phi$ near their thresholds.
A natural explanation is that it arises from $J/\Psi \to \gamma GG$,
where $G$ are gluons, followed by coupling of gluons to vector
mesons by colour neutralisation.
An essential feature of the $\gamma (\eta \pi \pi )$ data is the
requirement for a peak at $\sim 1600$ MeV, naturally associated
with $[\rho \rho ]_{L=1} \to [\eta \sigma]_{L=1}$ and
$[a_0(980)\pi]_{L=1}$.
The strategy adopted here is to examine how far one can get by fitting
this broad component by a single broad resonance $X$ in the
production process; exactly the same formula as (7) is used to fit
$\gamma (4\pi )$ data in the accompanying paper \cite {g4pi}.
The mass $M$ optimises empirically at $(2.04 \pm 0.10)$ GeV.
This is far removed from $\eta (1440)$ so the only significant phase
variation arises from (important) dispersive effects associated with
the opening of each threshold.
This broad background will be described as  `the broad $0^-$'.

Although Eq. (7) may look complicated, it is straightforward to
programme and gives an excellent fit to present data and
simultaneously to data on $J/\Psi \to \gamma 4\pi$.
Each of the channels $\eta \sigma$, $a_0(980)\pi$, $f_0(980)\pi$,
$K^*\bar K$, $\kappa \bar K$, $\rho \rho$, $\omega \omega$,
$K^* \bar K^*$ and $\phi \phi$ requires just one parameter
$g^2$ constrained (iteratively) to reproduce observed branching
fractions of these channels.
This constrains the $s$-dependence of the broad $0^-$ amplitude
closely.
The first five channels are essential to fit $\eta
\pi \pi$ and $K\bar K\pi$ data.
The inclusion of the $K^*\bar K^*$ threshold examines whether or not
the drop in the $\eta \pi \pi$ data above 1850 MeV can be explained by
the $K^*\bar K^*$ threshold.
The channels $\omega \omega$ and $\phi \phi$ have little
effect but are included for completeness.

Further details concern the form factor $F$ and centrifugal barriers
$B$ in Eq. (7).
The form factor for $J/\Psi \to X$, $X \to j$ is the product of a
weak form factor $F_p$ for the first step and a factor $F_d$ for
decays of $X$ through channel $j$.
Both are taken as Gaussians:
\begin {equation}
F = \exp (-\alpha _p K^2)\exp (-\alpha _d k^2) ,
\end {equation}
where $K$ is the production momentum of $X$ in the $J/\Psi $ rest frame
and $k$ is the decay momentum in the rest frame of X (for example,
the momentum of $\eta$ in the channel $\eta \sigma$).
Values of $\alpha _p$ and $\alpha _d$ are taken from the work on
$J/\Psi \to \gamma 4\pi$:
$\alpha _p = 0.06$ GeV$^{-2}$ and $\alpha _d = 2.25$ GeV$^{-2}$,
corresponding to RMS radii $R_p = 0.12$ fm, $R_d = 0.73$ fm.
The production form factor is very weak and has little effect here, but
is included for consistency with $\gamma (4\pi )$ data.

The factor $B$ is likewise the product of the $L' = 1$ centrifugal
barrier for $J/\Psi \to \gamma X$ and (where necessary) a centrifugal
barrier for the decay, for example an $L=1$ barrier for the decay of
$0^-$ to $K^*\bar K$.
As a reminder, for $L=1$,
\begin {equation}
B = kR/\sqrt {1 + k^2R^2};
\end {equation}
the same value of $R$ is used in the centrifugal barrier as in the form
factor, though this may be an approximation.

\section {Fits to data up to 1600 MeV}
Two types of fit will be discussed.
The first (Fit A in tables) tries to fit both $\eta \pi \pi$ and $K\bar
K\pi$ data with a single $\eta (1440)$ resonance.
The second (fit B) uses two separate resonances like $\eta (1405)$
and $\eta (1475)$, though their masses and widths need re-optimising.

\begin {figure}  [htb]
\begin {center}
\vskip -6mm
\epsfig{file=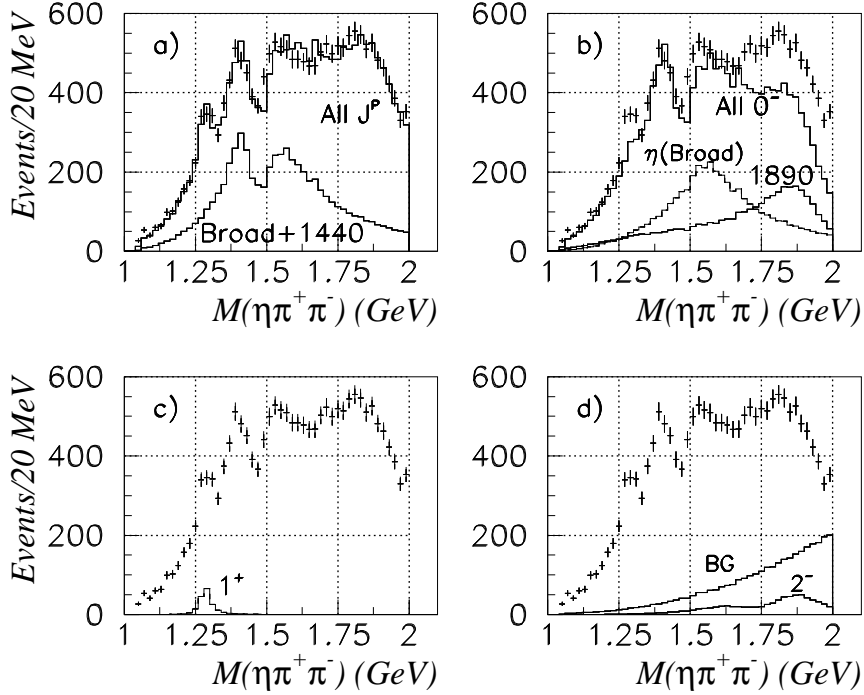,width=12cm}\
\vskip -6mm
\caption{The fit to $\eta \pi \pi$ data using $\eta (1440)$.
Points with errors show data.
In (a) the upper histogram shows the full fit;
the lower one shows the coherent sum of $\eta (1440)$ and
the broad $0^-$; (b) shows the total $J^P = 0^-$ contribution;
the lower histograms shows the broad
$0^-$ intensity from Eq. (7) and the $\eta (1890)$ contribution;
(c) the $1^+$ intensity; (d) experimental background (BG) and the $2^-$
contribution.}
\end {center}
\end {figure}

Fig. 7 shows details of the fit to $\gamma (\eta \pi \pi)$ data
with a single resonance which optimises at a mass of $1439 \pm 5$
MeV.
Panel (a) shows the overall fit to the mass distribution;
the lower histogram shows the coherent sum of $\eta (1440)$ and the
broad $0^-$.
Panel (b) shows the full $0^-$ intensity, also the broad component
from Eqs. (7)--(9) and thirdly $\eta (1890)$.
Panel (c) shows  the small peak due to $f_1(1285)$.
Panel (d) shows background as the upper histogram and the
small $2^-$ contribution required above 1600 MeV from
$\eta (1645)$ and $\eta (1870)$.
A detail is that it is necessary to check that the $f_0(980)\eta$
contribution to $\eta \pi \pi$ is consistent with data on $\eta
K^+K^-$.
The dashed curve on Fig. 1(f) shows the signal predicted for
$f_0(980)\eta$ with $f_0(980) \to K\bar K$.
It is consistent with data near 1600 MeV.

The sharp rise of the $\eta \pi \pi$ mass distribution at
1495 MeV is not fitted perfectly in Fig. 7(a): the data would prefer
a rise 10 MeV lower than the fit.
The rapid rise requires a strong contribution from $f_0(980)\eta$.
The fit which is shown uses the BES parametrisation of $f_0(980)$
\cite {phipp}.
A reparametrisation is desirable including the threshold cusp in the
real part of the amplitude at the $K\bar K$ threshold.
This has been tried, but does not cure the problem: it simply sharpens
the $f_0(980)$ peak near the $K\bar K$ threshold.

\begin {figure}  [htb]
\begin {center}
\vskip -10mm
\epsfig{file=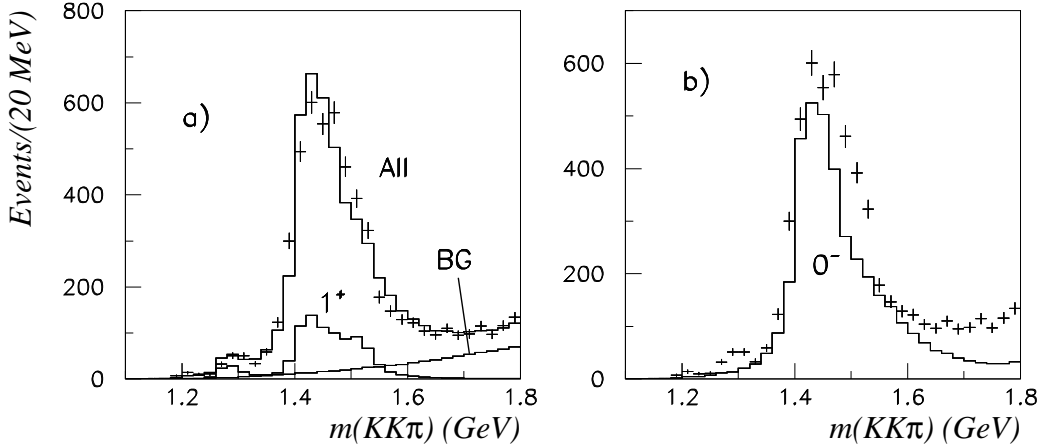,width=14cm}\
\vskip -6mm
\caption{The fit to $K\bar K\pi$ data.
In (a) the upper histogram shows the full fit and the lower one the
$J^P = 1^+$ contribution from $f_1(1285)$,$f_1(1420)$ and
$f_1(1510)$; the experimental background is labelled BG.
(b) shows the $0^-$ contribution alone.  }
\end {center}
\end {figure}

Fig. 8 shows the fit to  $\gamma (K\bar K\pi )$ data.
The peak sits on a small experimental background shown in (a)
and a slowly rising physics background, clearly visible
above experimental background.
This physics background has $J^P = 0^-$ and is parametrised by the
broad $0^-$ contribution decaying to $K^*\bar K$, $\kappa \bar K$ and
$a_0\pi$,  $a_0 \to K\bar K$.
The coupling constants of the first two are fitted freely; the last
component is derived from the $a_0(980)$ fitted to
$\eta \pi \pi$ data using the branching ratio between
$\eta \pi$ and $K\bar K$ in Ref. \cite {a0980}.

The $K\bar K\pi$ peak of Fig. 8 is decidedly asymmetric.
The main source of the asymmetry can immediately be traced to the
$s$-dependence of the $K^*\bar K$ decay channel whose contribution
is shown in Fig. 8(b).
The lower histogram in Fig. 8(a) shows the contribution from
$J^P = 1^+$.
All fits require some $1^+$ contribution above $f_1(1420)$;
this extra contribution is parametrised in the present fit by
$f_1(1510)$ with mass and width quoted by the PDG.
Changing the parameters to values of Aston et al. \cite {fA1510}
changes log likelihood only by 3.6.

Attempts have been made to fit the small shoulder
in Fig. 8 at 1530 MeV in several ways.
This mass is close to the opening of the $K\bar K\eta$ channel at 1539
MeV.
It raises  the possibility of a drop in the line-shape associated
with the opening of this channel; this is a unitarity effect - the
opening of the $\eta K\bar K$ channel robs $K\bar K\pi$ of some
signal.
However, this is immediately excluded by the small branching ratio
of $\eta K\bar K$ compared to $\eta \pi \pi$.
A more subtle possibility is that it may be associated with the
$f_0(980) \eta$ channel, which also peaks at 1539 MeV, since $f_0(980)$
itself peaks at the $K\bar K$ threshold.
The fit to $\eta \pi \pi$ data requires a strong contribution from
$f_0(980)\eta$.
However, it is spread out quite significantly according to phase space
for the $f_0(980)\eta$ final state, as shown below in Fig. 13(a);
that is too broad to explain a narrow shoulder at 1530 MeV.
Note that there is independent evidence for this $f_0(980)\eta$
contribution to $J^P = 0^-$ \cite {Anisovich}.

The $f_2(1565) \to [a_2(1320)\pi ]_{L=1}$ does not account
for the 1550 MeV structure in $\eta \pi \pi$ or the 1530 MeV
shoulder in $K\bar K\pi$;
there is no visible $a_2(1320)$ peak at this mass in $\eta \pi$ or
$K\bar K$ and the angular dependence is wrong.
Furthermore, the magnitude which can be fitted to $f_2(1565)$ in
$\gamma 4\pi$ rules out the possibility of explaining the 1530 MeV
shoulder with decays to $[a_2(1320)\pi]_{L=1}$.

\begin {figure}  [htb]
\begin {center}
\vskip -10mm
\epsfig{file=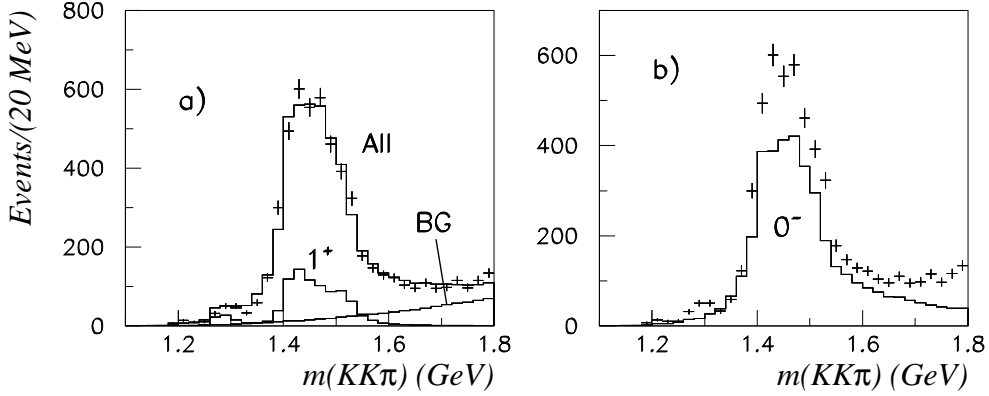,width=14cm}\
\vskip -6mm
\caption{As Fig. 8 after the inclusion of $\eta (1565)$.}
\end {center}
\end {figure}

One further possible explanation remains.
In the accompanying paper on $J/\Psi \to \gamma 4\pi$, there is evidence
that the $\rho \rho$ peak at $\sim 1600 $ MeV is actually resonant at
$\sim 1560$ MeV.
If this possible resonance has a weak decay to $K^*\bar K$,
the shoulder at 1530 MeV can be fitted as an interference effect.
The resulting fit is shown in Fig. 9 in identical format to Fig. 8;
the addition of $\eta (1560)$ with a width of 280 MeV improves log
likelihood by 62.
This explanation is clearly speculative.
It is to be expected that two neighbouring resonances at 1440 and
1560 MeV will provide considerably more flexibility in the fit than
$\eta (1440)$ alone.
However, the magnitude of the contribution from $\eta (1560)$ is
fairly small, as would be expected if its mixing with $s\bar s$ is
small; its intensity is $19\%$ of that from $\eta (1440)$.
This additional resonance does not provide the 10 MeV mass shift
near 1500 MeV needed for a perfect fit to the $\eta \pi \pi$ mass
projection of Fig. 7(a).
A possible remedy for this detail will be discussed in Section 7.

\begin {figure}  [htb]
\begin {center}
\vskip -14mm
\epsfig{file=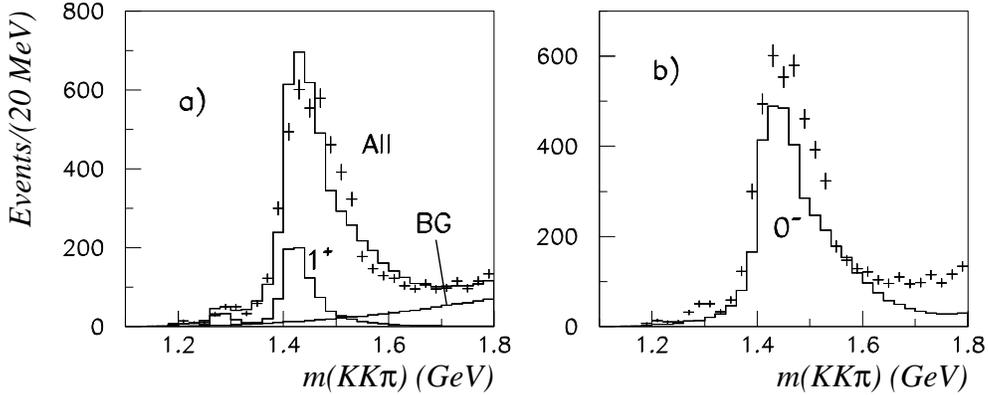,width=14cm}\
\vskip -6mm
\caption{The fit to $K\bar K\pi$ data without
$f_1(1510)$.
In (a) the upper histogram shows the full fit and the lower one the
$J^P = 1^+$ contribution from $f_1(1285)$ and  $f_1(1420)$;
(b) shows the $0^-$ contribution alone.  }
\end {center}
\end {figure}

Let us now turn to the $J^P=1^+$ contribution.
Fig. 10(a) shows the poor fit without $f_1(1510)$.
The remaining component due to $f_1(1285)$ and $f_1(1420)$ is shown
as the lower histogram.
The $f_1(1420)$ is parametrised  including the
$s$-dependence of the $K^*\bar K$ channel and the associated
dispersive term.
The result is to extend the tail of $f_1(1420)$ to
slightly higher masses, but does not substitute for $f_1(1510)$.
The mass of $f_1(1420)$ is held fixed at the PDG value and its
full width at half maximum is fixed to the width quoted by the PDG.
The magnitude fitted to $f_1(1420)$ increases in Fig. 10 to 743
events compared with 488 in Fig. 8.
The reason for this is the requirement for some significant $1^+$ signal
in the mass range around 1500 MeV.
An alternative is to increase the width of $f_1(1420)$ quite
significantly.
This possibility will be discussed in Section 7.

\subsection {The fit with separate $\eta (1405)$ and $\eta (1475)$}
The full $s$-dependence is fitted to both states including the
dispersive terms.
This requires a determination of the channels which contribute to
each.

The $a_0(980)\pi$ and $\eta \sigma$ channels are naturally
attributed to $\eta (1405)$, as in earlier work.
Attempts to include these channels into $\eta (1475)$ lead to
negligible contributions.
The $K^*\bar K$ channel is naturally associated with $\eta (1475)$,
as in earlier work.
If it is added to $\eta (1405)$, log likelihood improves only by
3.9, which is insignificant.
Here, it is necessary to comment on earlier work of Adams et al.
\cite {Adams}.
Their Fig. 4(d) shows a low mass $K^*\bar K$ component in addition to
that of $\eta (1475)$.
However, from the present analysis it is clear that this is an artefact
arising from their free fit to $a_0(980)\pi$, $a_0 \to K\bar K$.
The resulting large $a_0(980)$ intensity in $K\bar K
\pi$ is inconsistent with the $\eta \pi \pi$ data by a factor 10, so a
constraint on the magnitude of $a_0(980) \to K\bar K$ is essential.
This eliminates any $K^*\bar K$ contribution from $\eta (1405)$.

If $\kappa \bar K$ is fitted freely to both $\eta (1405)$ and $\eta
(1475)$, it goes almost entirely into $\eta (1405)$.
Removing the small component from $\eta (1475)$ makes log likelihood
worse by only 6.2, which is barely significant.
Furthermore, it removes a destructive
interference with $K^*\bar K$, symptomatic of instability.
The final fit includes $\kappa \bar K$ only in $\eta (1405)$.

The remaining component is $f_0(980)\eta$.
It improves log likelihood of the fit by 162, a
decisive amount.
It plays a strong role in creating the sharp rise in the
$\eta \pi \pi$ mass projection of Fig. 7(a) at 1500 MeV.
One might therefore naturally associate it with $\eta (1475)$.
However, the fit contradicts this possibility.
A free fit to both $\eta (1405)$ and $\eta (1475)$ leads to negligible
contribution to $\eta (1475)$ from $f_0(980)\eta$.
It is easy to locate the reason.
Associated with the $f_0(980)\eta$ channel is a large dispersive
term $m(s)$ in the denominator of any resonance to which is couples.
The phase variation created by this dispersive term is not compatible
with $\eta (1475) \to f_0(980)\eta$.
The final outcome is that $\eta (1475)$ has only a $K^*\bar K$
contribution and all other channels contribute via $\eta (1405)$.

The peak in  $\eta \pi \pi$ data is at $1415 \pm 5$ MeV.
However, because of the interference with the broad $0^-$
contribution, the mass which can be fitted is quite flexible,
with an uncertainty of $\pm 20$ MeV.
The fit actually optimises at 1429 MeV because the phase variation
associated with $m(s)$ in the Breit-Wigner denominator favours the
higher mass.

For $\eta (1475)$, there is a strong correlation between the
fitted width and the optimum mass.
The strict optimum is at 1505 MeV.
However, log likelihood is worse by only 4 if the PDG mass of
1476 MeV is used.
In order to allow comparison with published results, the mass is
fixed at the PDG average, 1476 MeV and the width is then optimised.
The fitted width at half-maximum is 78 MeV, compared with the
PDG average of $85 \pm 9$ MeV.
The fit to the $\eta \pi \pi$ mass distribution using these values is
shown in Fig. 11.
Near the 1415 MeV peak, it is somewhat different in shape to that of
Fig. 7.
The fit to the rapid rise at 1500 MeV is no better than in Fig. 7(a).

\begin {figure}  [htb]
\begin {center}
\vskip -6mm
\epsfig{file=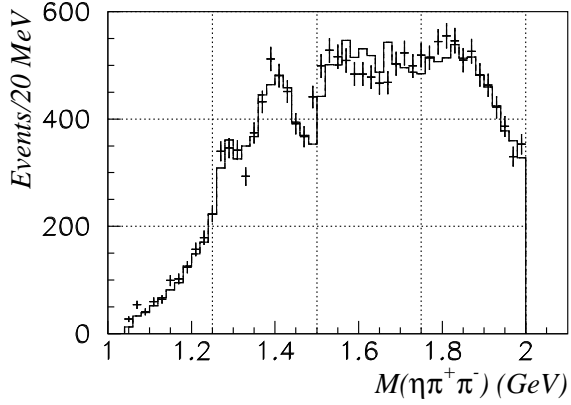,width=8cm}\
\vskip -6mm
\caption{The fit to the $\eta \pi \pi$ mass distribution
using both $\eta (1405)$  and $\eta (1475)$.  }
\end {center}
\end {figure}

The fit to $K\bar K\pi$ data is shown in Fig. 12.
The mass distribution of the peak is fitted rather better than in
Fig. 8.
The reason is obvious: two resonances allow much greater flexibility.
In particular, the phase variation with $s$ allowed by 2 resonances
has much greater freedom than the fixed relation between magnitude and
phase for $\eta (1440)$ alone.
With two resonances, it is possible to
arrange that the mass distribution in Fig. 12 rises rapidly and falls
rapidly, but with something approaching a flat top.

However, there is very little difference in log likelihood between
fits with one resonance or two.
The fit to $\eta \pi \pi$ data with two states is better by 6.5 in log
likelihood and the fit to $K\bar K\pi$ data is better by only 2.6,
making a total of 9.1 for the two sets of data.
This hardly warrants confidence in the need for two resonances.

\begin {figure}  [htb]
\begin {center}
\vskip -12mm
\epsfig{file=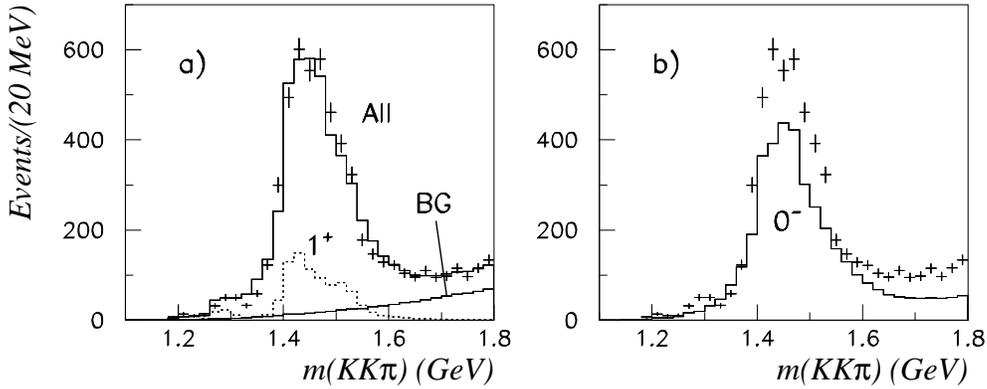,width=14cm}\
\vskip -6mm
\caption{The fit to the $K\bar K\pi$ mass distribution
using both $\eta (1405)$  and $\eta (1475)$.  }
\end {center}
\end {figure}

Table 1 shows changes in log likelihood for both types of fit when
individual components are removed and magnitudes and phases of all
other components are re-optimised.
The third column shows results for fit A with one resonance,
$\eta (1440)$, and the fourth shows results for fit B with two
resonances.
This Table also includes branching fractions for channels or
combinations of channels (where interferences are large).
A warning is that for $\eta (1440)$ there are strong interferences
between $\kappa \bar K$ and $K^* \bar K$ and these are somewhat
different for fits A and B.
There are likewise large interferences between $\kappa \bar K$
and $K^*\bar K$ for the broad $0^-$ signal.
Errors given for branching fractions cover uncertainties in
interferences.
Because of large interferences between the broad $0^-$ and
$\eta (1890)$, branching fractions of Table 1 do not add up to the
total branching fraction given in Eq. (5).

The one feature which might favour a fit with two resonances rather
than one is the $f_1(1510)$ contribution.
Its parameters are fixed at PDG averages: $M = 1518$ MeV,
$\Gamma = 73$ MeV.
It improves the fit with a single $\eta (1440)$ and an $f_1(1510)$,
which is regarded by some authors as questionable, by
54.0.
The improvement for the fit with 2 resonances is 38.9.
So the fit without any $f_1(1510)$ is better for fit B.
Both fits require some $f_1(1510)$ contribution, or something
resembling it.
A  test has been made of the requirement for $f_1(1510)$, as distinct from
a $0^-$ contribution.
A further $0^-$ state is added to fit B using
$\eta (1405)$ and $\eta (1475)$.
It is given identical mass and width to $f_1(1510)$.
The result is that the $f_1(1510)$ contribution
dominates over that from $\eta (1510)$, confirming a definite
requirement for a $1^+$ component in this mass range.
In Section 7, a possible explanation in terms of the line-shape
of $f_1(1420)$ will be presented.

\begin{table}[htb]
\begin {center}
\begin{tabular}{cccccc}
\hline
Resonance          & Channel                & Fit A & Fit B
& Br. Fr. (A) &Br. Fr. (B) \\\hline
$\eta (1440)$ or   & $a_0\pi$,$a_0 \to K\bar K$  & 8.7   &  8.1
& 0.1 & 0.15 \\
$\eta (1405+1475)$ &$\kappa \bar K$ & 131   & 80
& $2.1 \pm 0.3$ & $3.3 \pm 0.5$ \\
                   &$K^*\bar K$     & 441   & 382
& $ 2.5 \pm 0.4$ & $3.8 \pm 0.6$ \\
                   & $a_0\pi$,$a_0 \to \eta \pi$ & 47 & 55
& $0.40 \pm 0.15$ & $ 0.6 \pm 0.2$ \\
                   & $\eta \sigma$    & 39    & 81
& $0.3 \pm 0.1$ & $0.4 \pm 0.1$ \\
                   & $f_0(980)\eta$ & 151   & 162
& $ 1.0 \pm 0.3$ & $1.5 \pm 0.6$ \\\hline
Broad $0^-$        & $a_0\pi$,$a_0 \to K\bar K$   & 10    & 10 \\
                   &$\kappa \bar K$ & 147   & 34
& $1.6 \pm 0.7$ & $1.3 \pm 0.4$ \\
                   &$K^*\bar K$     & 132   & 111
& $1.8 \pm 0.5$ & $3.3 \pm 1.0$ \\
                   &$a_0\pi$,$a_0 \to \eta \pi$ & 115 & 123\\
                   &$\eta \sigma $  &  80   & 141  \\
& $\eta \sigma + a_0\pi \to \eta \pi \pi$ & & &
$18.4 \pm 9.7$ & $19.0 \pm 10.0$ \\\hline
$\eta (1890)$      &$a_0\pi$,$a_0 \to K\bar K$ & 16 & 21 \\
                   &$a_0\pi$,$a_0 \to \eta \pi$ & 177 & 197 \\
                   &$\eta \sigma$   & 134 & 157   \\
                   &$f_0(980)\eta$  & 81    & 107
&$1.2 \pm 0.2$ & $1.4 \pm 0.2$\\
& $\eta \sigma + a_0\pi \to \eta \pi \pi$ & & &
$15.3 \pm 9.7$ & $16.4 \pm 10.0$ \\\hline
$f_1(1285)$        &$\kappa \bar K$ & 70    & 143
& $0.18 \pm 0.03$ & $0.14 \pm 0.03$ \\
                   &$a_0 \pi$,$a_0 \to \eta \pi$& 92    & 105
& $0.9 \pm 0.3$ & $0.9 \pm 0.3$ \\
                   &$\eta \sigma $  &  9    & 11  \\
$f_1(1420)$        &$K^*\bar K$     & 129   & 152
& $0.6 \pm 0.1$ & $1.3 \pm 0.2$ \\
$f_1(1510)$        &$K^*\bar K$     & 54    & 39
& $0.4 \pm 0.1$ & -  \\
$\eta _2(1645)$    & all $\eta \pi \pi$ & 11& 11
& $0.10 \pm 0.03 $ & $0.10 \pm 0.03$ \\
$\eta _2(1870)$    & all $\eta \pi \pi$ & 83& 84
& $1.1 \pm 0.3$ & $1.1 \pm 0.3$  \\\hline
\end{tabular}
\caption{Changes in log likelihood
when individual components are removed and all other components
are re-optimised in magnitude and phase.
Column 3 shows results fitting $\eta (1440)$ alone and column 4
results fitting both $\eta (1405)$ and $\eta (1475)$.
Columns 5 and 6 show branching fractions for $J/\Psi \to \gamma + X$,
multiplied by $10^4$.}
\end{center}
\end{table}

\begin {figure}  [htb]
\begin {center}
\vskip -10mm
\epsfig{file=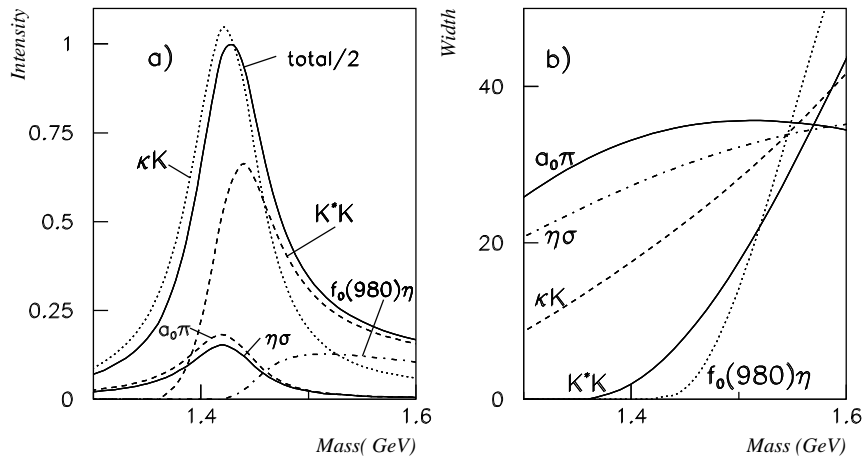,width=13cm}\
\vskip -6mm
\caption{(a) Contributions to the line-shape of $\eta (1440)$, Fit A;
(b) the variation with mass of phase space for each channel.
The $a_0\pi$ signal refers to $\eta \pi \pi$; that for $a_0 \to K\bar
K$ is 24\% of the signal in $\eta \pi \pi$. }
\end {center}
\end  {figure}

\subsection {Further details of the fit with $\eta (1440)$}
Fig. 13(a) shows as the full curve the overall line-shape of
$\eta (1440)$, scaled down by a factor 2 in order to display other
contributions.
The peak for $K^*\bar K$ is at 1440 MeV, but there is a pronounced
tail at high mass.
One should note that in Obelix and Crystal Barrel data, this high
mass tail is suppressed strongly by the phase space for production
in $\bar pp$ annihilation at rest.

From Fig. 13(a) it is quite difficult to decide on the
centre of gravity of the $K^*\bar K$ contribution.
Averaging this contribution up to 1600 MeV it is at 1474 MeV,
remarkably close to the PDG mass for $\eta (1475)$, namely 1476
MeV.
Putting this in reverse, if one unfolds the expected $s$-dependence
of $K^*\bar K$, the resonance  mass should be expected at 1440 MeV, and
agrees with the fit with a single resonance.
The $f_0(980)$ channel has a weak maximum near 1530 MeV, but
also extends strongly to high mass.
At high mass there are strong interferences between $K^*\bar K$ and
$\kappa \bar K$.
Constructive interferences between them shifts the peak in $K\bar
K\pi$ up to 1450 MeV.
Fig. 13(b) shows the dependence of phase space for each channel on mass,
but with an arbitrary scale for each channel, so as to fit all the
curves on to one figure.

\begin {figure}  [htb]
\begin {center}
\vskip -12mm
\epsfig{file=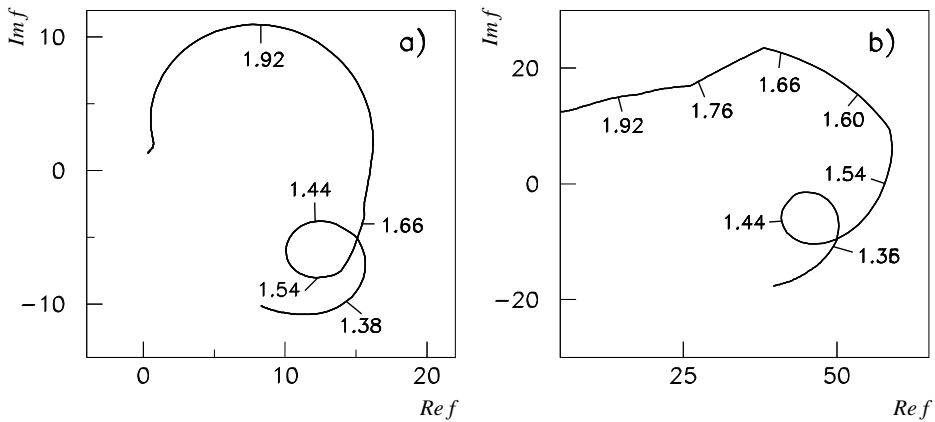,width=14cm}\
\vskip -6mm
\caption{Argand diagrams for (a) $a_0(980)\pi$ and (b)
$\eta \sigma$ for the coherent sum of $\eta (1440)$ and the broad $0^-$
contribution. Masses are marked in GeV.}
\end {center}
\end {figure}

Fig. 14(a) shows the Argand diagram for the coherent sum of
$a_0(980)\pi$ amplitudes from $\eta (1440)$ and the broad $0^-$ signal
on which it sits.
The diagram is drawn for an $a_0(980)$ mass of 980 MeV, so as to avoid
the effect of phase space on the integrated amplitude.
Numbers on the figure indicate $\eta \pi \pi$ masses in GeV.
Fig. 14(b) shows the Argand diagram for the $\eta \sigma$ amplitude
evaluated for a $\sigma$ mass of 800 MeV.
Up to 1415 MeV, the real parts of the amplitudes are large.
The dispersive contribution to the real part is large because the
amplitude senses the steep rise at higher masses in the imaginary part
- an impending barrier, which gets closer as the mass rises.
As the $K^*\bar K$ channel opens, there are two effects.
Firstly, it increases the magnitude of the Breit-Wigner denominator and
damps all $\eta \pi \pi$ amplitudes.
Secondly, the real part of the amplitude is approximately proportional
to the gradient of the imaginary part.
It turns over rapidly at 1420 MeV and plunges through zero on resonance
at 1440 MeV; these two features produce the narrow peak at 1415 MeV
in $\eta \pi \pi$.
The $f_0(980)\pi$ amplitude contributes strongly to the steep rise in
$\eta \pi \pi$ at 1500 MeV.
The cusp at 1600 MeV in the broad $0^-$ contribution is largely
responsible for the shoulder in $\eta \pi \pi$ data at 1550 MeV.

Table 2 gives branching ratios integrated from threshold up to 1.8
GeV.
Systematic errors from this cut-off are a factor $(1.0 \pm 0.1)$ if the
cut-off is reduced to 1650 MeV or increased to 2100 MeV.
Errors cover variations observed in many alternative fits; they also
cover the (small) discrepancies between the fits with one resonance or
two.
Note that there is large interference between $a_0(980)\pi $ and $\eta
\sigma$.

\begin{table}[htb]
\begin {center}
\begin{tabular}{cccc}
\hline
Channel                     & Branching fraction ($\%$)\\\hline
$K^*\bar K$                 & $40 \pm 5$ \\
$\kappa \bar K$             & $35 \pm 5$ \\
$\eta \sigma$               & $4.1 \pm 1.6$ \\
$a_0\pi$,$a_0 \to \eta \pi$ & $5.5 \pm 2.1$ \\
$a_0\pi$,$a_0 \to K\bar K$       & $1.3 \pm 0.5$  \\
$f_0(980)\eta$              & $14 \pm 3$  \\\hline
\end{tabular}
\caption{Branching fractions of $\eta (1440)$.}
\end{center}
\end{table}

\subsection {The branching ratio $K\bar K\pi/\eta \pi \pi$}
From Table 2, the ratio
\begin {equation}
\frac {BR(\eta (1440) \to K\bar K\pi)}
{BR(\eta (1440) \to \eta \pi \pi)} = 3.0 ^{+1.5}_{-0.9}.
\end {equation}
The errors have been obtained by adding in quadrature the errors for
$K^*\bar K$ and $\kappa \pi$ to get overall branching fractions for
$K\bar K\pi$ in the range $(82-68)\%$, then dividing by the
corresponding $\eta \pi \pi$ fraction.

In their review, Masoni et al. \cite {Masoni} present in their
Table 8 their estimate of the same branching ratio, or more
exactly its inverse.
Converting their entries to the definition used here, they estimate
values in the range $0.98 \pm 0.37$ from Mark III and DM2 data,
$0.30 \pm 0.1$ using Crystal Barrel and bubble chamber data of
Baillon et al. \cite {Baillon}, or $0.11 \pm 0.05$ from
Crystal Barrel and Obelix data.
They use the disagreement between these sources as an argument in
favour of separate $\eta (1405)$ and $\eta (1475)$.

However, this analysis is misleading.
When one analyses data for  $\bar pp $ annihilation,
it is vital to allow for the suppression of the $K^*\bar K$ channel
by the available phase space.
This was ignored by Masoni et al.
The factor $H(m)$ multiplying the line-shape of $\eta (1440)$ is:
\begin {equation}
H(m) = \int ^{(2M_p - \sqrt {s_1})^2}_{4m^2_{\pi}} ds_1 \,
\frac {|p_{prodn}|}{2M_p}
\frac {|p_1(\sigma \to\pi \pi )|}{\sqrt {s_1}} |T_\sigma (s_1)|^2
|F(p_{prodn})|^2.
\end {equation}
Here $s_1$ is the mass squared of the spectator $\pi \pi$ pair, and
$p_{prodn}$ is the momentum of $\eta (1440)$ or the spectator
$\sigma$ in the $\bar pp$ rest frame.
Next, $M_p$ and $m_\pi$ are the masses
of the proton and pion.
Also  $T_\sigma$ is the scattering amplitude for the
spectator $\pi \pi$ S-wave pair and $F$ is the form
factor for the production process, taken here as
$\exp (-2.25 p_{prodn})$ with $p_{prodn}$ in GeV/c.

\begin {figure}
\begin {center}
\vskip -6mm
\epsfig{file=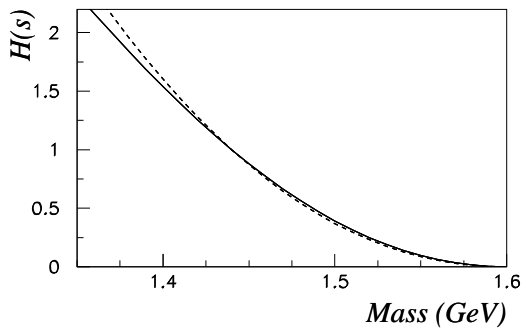,width=8cm}\
\vskip -6mm
\caption{The factor $H(m)$ allowing for the phase space and
dynamics of the production process $\bar pp \to \eta (1440) \sigma$
as a function of the mass $m$ of $\eta (1440)$.
The full curve shows production via the $\sigma$ pole for the spectator
pair and the dashed curve the result using the $\pi \pi$ elastic
scattering amplitude.
It is normalised to 1 at 1440 MeV.}
\end {center}
\end {figure}
The variation of $H(m)$ with the mass of $\eta (1440)$ is shown in
Fig. 15.
It is not clear whether the $\pi \pi$ spectator pair will be produced
via the $\sigma$ pole or via the $\pi \pi$ elastic scattering amplitude.
The former is shown as the full curve and the latter as the dashed
curve.
They are sufficiently close that one can work with the average.
The high mass tail of $\eta (1440)$ is cut off strongly in $\bar pp$
annihilation.
It goes to zero at 1.60 GeV: the highest mass which can be
produced in $\bar pp$ annihilation with two spectator pions.

There are two consequences.
Firstly, the factor $H(m)$ amplifies the lower side of $\eta (1440)$
with respect to the upper side.
For the narrow $\eta \sigma$ and $a_0(980)$ channels, the effect
is to shift the apparent peak of $\eta (1440)$ downwards by
4.7 MeV; i.e. the Breit-Wigner mass is actually 4.7 MeV higher than
is fitted with a Breit-Wigner amplitude of constant width.
For the $\kappa K$ channel, the effect is larger because the phase
space for this channel, shown on Fig. 13(b), increases significantly
with $m$.
The result is that the mass fitted to this channel is moved downwards
by 9.7 MeV, so the true mass fitted as a Breit-Wigner resonance of
constant width should be increased by this amount.

The second effect is that a considerable part of the $K\bar K\pi$
signal is lost in data on $\bar pp$ annihilation; all of it
is lost above 1.60 GeV.

A check will be presented here on the ratio $K\bar K\pi /\eta \pi \pi$
from data on $\bar pp$ annihilation.
Attention will be drawn to many sources of uncertainty.
The check is done using data of Anisovich et al. \cite {Nana} for
$\bar pp \to (\eta \pi ^+\pi ^-)(\pi ^+\pi ^-)$ and bubble chamber data
of Baillon et al \cite {Baillon}.
Table 2 of Anisovich et al. gives branching fractions for all
amplitudes fitted to their data, in particular $\eta
(1440) \to \eta \sigma$ and $a_0(980)\pi$.
This Table reveals an important point.
There is large destructive interference in those data between these
two amplitudes.
This destructive interference has a strong effect on the
branching fraction quoted for $\eta (1440)$.
However, in reality there are major uncertainties in determining
the interference.
There are actually four charge cominations in those data
allowing considerable freedom in interferences.
The individual contribution from $\eta \sigma$ is quoted as
$2.31 \times 10^{-3}$, that from $a_0(980)\pi$ is $9.6 \times 10^{-3}$
and the interference between them contributes $-2.28 \times 10^{-3}$.
It clearly makes a large difference how this interference is taken into
account.

Relative magnitudes of all decay channels of a resonance should be the
same in all production processes.
The procedure adopted by Anisovich et al., and used here, is to derive
coupling constants from the fit to data, then evaluate the intensity of
amplitudes by taking the modulus squared of individual amplitudes.
Interference between decay channels of one resonance are kept, since
these are a property of the resonance.
Interferences with other resonances in the data are ignored because
such interferences are a feature of the particular environment in which
the $\eta (1440)$ is produced.

The resulting branching fraction for production of $\eta (1440)$
in $\eta \pi \pi$ via $\bar pp$ annihilation is $1.09 \times 10^{-3}$
in the observed charge states.
This needs to be multiplied by a factor $9/4$ to allow for other
charge states, giving a result of $2.45 \times 10^{-3}$ for
$\bar pp \to \eta(1440)$, $\eta (1440) \to \eta \pi \pi$.
The result for all charges of $K\bar K\pi$ quoted by Baillon et al.
is $(2.0 \pm 0.2) \times 10^{-3}$.
This gives a ratio $R_{1440} =
BR(\eta (1440) \to K\bar K\pi)/BR(\eta (1440) \to \eta \pi \pi
= 0.82 $.
Masoni et al. use the smaller branching ratio
$0.61 \pm 0.19$ found by Amsler et al. \cite {Amsler} in an earlier
analysis of $\bar pp$ data.
However, that publication does not specify if any allowance is
made for the effect of destructive interference between
$\eta (1440) \to \eta \sigma$ and $a_0(980)\pi$.

The effect of $H(m)$ on the integrated intensities for final states
$\eta \sigma$ and $a_0(980)\pi$ is small because the phase space
$\rho (s)$ for these channels varies slowly with $m$ and there is an
approximate cancellation between upper and lower sides of the
resonance.
For the $\kappa \bar K$ channel, there is a noticeable effect  because
of its fairly rapidly increasing phase space, see Fig. 13(b).
Using Fig. 15, the $\kappa \bar K$ component in $\bar pp$ annihilation
needs to be multiplied by a factor 1.38 and the $K^*\bar K$
contribution needs to be multiplied by $\sim 3.1$.
If we take the relative contributions of $K^* \bar K$ and
$\kappa \bar K$ to be in the ratio 40:35 of Table 2, the
result is to increase the $K\bar K\pi$ branching fraction of
Baillon et al by a factor 2.3, resulting in a value $R_{1440}$ of 1.88.

This is still a little smaller than the value derived from Table 2,
namely $3.0 ^{+1.5}_{-0.9}$.
However, the essential conclusion is that the branching fraction
between $K\bar K\pi$ and $\eta \pi \pi$ is subject to a large
systematic correction when evaluated from $\bar pp$ data.
The result from $J/\Psi $ data does not suffer from these systematic
problems.
Note that the line-shapes fitted to present data are perturbed by the
slowly varying factor $P^3_\gamma$ for production with $L'=1$; this
factor and the associated form factor have been included in the fit
and the arithmetic of branching ratios.
One should also note that the signal assigned by Obelix to the
$K\bar K\pi$ channel is not visible as a peak in the data, but is
obtained from the difference in the line-shape of the $K\bar
K\pi$ signal from a Breit-Wigner of constant width fitted to $\eta
(1405)$.
A further detail is that the phases of individual decay channels are
affected by multiple scattering with other components of the fit.
They may differ from one production process to another.
As a result, it is difficult to give a definitive result for the
branching ratio between $K\bar K\pi$ and $\eta \pi \pi$.

The reality of the situation is that there is major flexibility in
the magnitudes of contribution  which can be fitted to $\eta (1440)$
in $\eta \pi \pi$ channels.
In the data presented here, there is flexibility in what can be fitted
to the broad interfering component.
This flexibility is apparent from the fact that fits with a single
$\eta (1440)$ or separate $\eta (1410)$ and $\eta (1475)$ differ
very little in log likelihood.
There is yet further flexibility if the $\eta \sigma$ channel receives
additional contributions proportional to the $\pi \pi$ elastic
amplitude.

\section {Fits to data above 1600 MeV}
\begin {figure}
\begin {center}
\epsfig{file=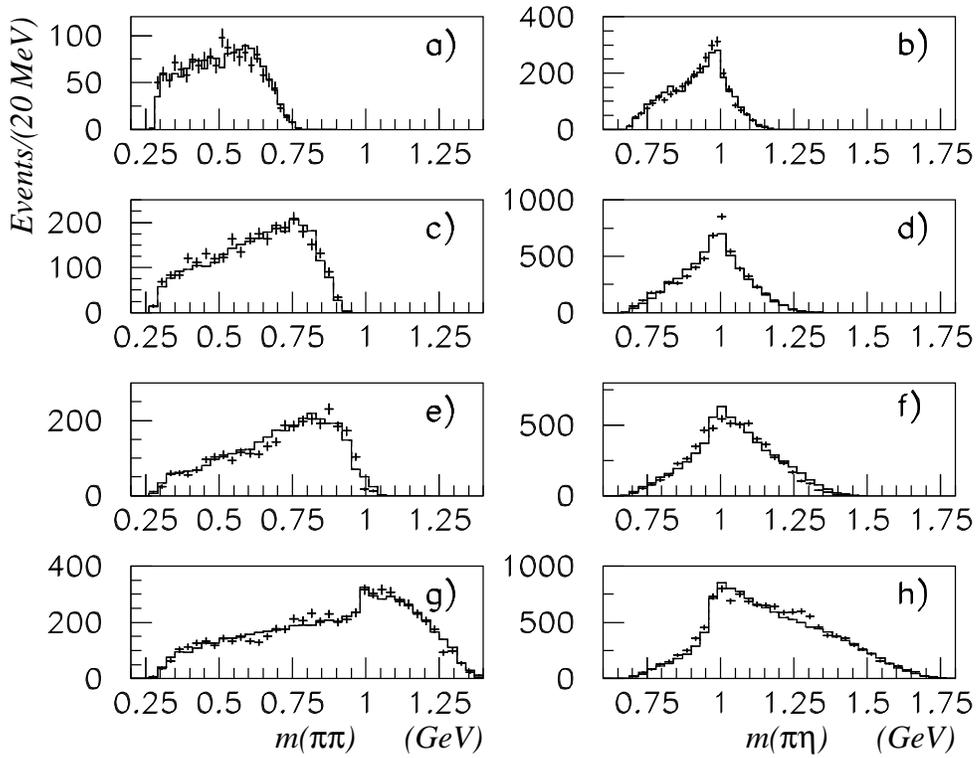,width=15cm}\
\vskip -16mm
\caption{Fits to mass projections on to
(a) $M(\pi \pi )$ and (b) $M(\pi \eta)$ for the $\eta \pi \pi$ mass
range 1230--1330 MeV, compared with the fit (histogram); (c)--(d) the
mass range 1350--1480 MeV; (e)--(f) 1480--1600 MeV; (g)--(h) 1700--1900
MeV. }
\end {center}
\end {figure}

In $\eta \pi \pi$ data shown in Fig. 7, the broad $0^-$ signal
in Fig. 7(b) obviously fails to explain fully the $\eta \pi \pi$
signal above 1600 MeV.
The $K^*\bar K^*$ threshold may contribute something to the drop
above 1870 MeV, but alone does not fit the high mass spectrum.
The remedy adopted here is to fit a resonance form as an expedient,
although there is no phase information, hence no definite evidence for
a resonance.
The optimum fitted mass and width are $1890 \pm 20$ MeV, $\Gamma =
260 \pm 35$ MeV.
The strongest decays are to $a_0(980)\pi$, but there is also a
rather strong decay to $f_0(980)\eta$.
Fig. 16 show mass projections from $\eta \pi \pi$ Dalitz plots
on to $\pi \pi$ and $\eta \pi$ mass.
The $f_0(980)$ is clearly visible in Fig. 16(g) at 1 GeV.
In Figs. 16(h), there is a a small but definite $a_2(1320)$ peak.
It does not fit naturally to decays of $\eta _2(1870)$ and/or $\eta
_2(1645)$.
It is of roughly the right magnitude to originate from
$a_2(1320)\rho$ background.
The $\rho(770)$ signal from this background
is also just visible in Fig. 16(g).

On the upper side of the 1850 MeV peak is a small $2^-$ component,
which may be fitted with $\eta _2(1870)$.
When this contribution is omitted from the fit, log likelihood gets
worse by 84.
The relative intensities in $a_0(980)\pi$,
$a_2(1320)\pi$, $f_2(1270)\eta$ and $\sigma \eta$ decays are fixed to
accurate values from Crystal Barrel data \cite {CBAR}.
Production phases,  attached to each amplitude in the isobar
model, are fitted freely.
So there are five fitted parameters.
Statistically, the significance of the $2^-$ component is 10 standard
deviations, despite the signal being small.
The reason for this significance is that amplitudes have a
distinctive angular dependence which is readily identified.
For completeness, a contribution is also included for $\eta _2(1645)$.
It is much smaller and barely significant.
In Ref. \cite {CBAR}, it is suggested that $\eta _2(1645)$ is a standard
$q\bar q$ state, the isospin partner to $\pi _2(1670)$; $\eta _2(1870)$
is interpreted as a hybrid.
This interpretation is consistent with the observations here.
The production of a hybrid is naturally explained in terms of
$c\bar c \to \gamma gg$, followed by conversion of one gluon
$g$ to $q\bar q$.

\subsection {The magnitude possible for $X(1835) \to \eta \pi \pi$}
The BES Collaboration has presented evidence for an $X(1835)$ in
$\eta '\pi \pi$ with a production branching fraction
$BR(J/\Psi \to \gamma X).BR(X \to \pi ^+\pi ^- \eta ') =
[2.2 \pm 0.4(stat) \pm 0.4(syst)] \times 10 ^{-4}$
and a width of 68 MeV \cite {X1835}.
There is no obvious signal in $\eta \pi ^+\pi ^-$ data of Fig. 7,
although the fit to data is slightly ragged in this mass range.

A free fit including it with published mass and width gives
189 $f_0(980)\pi$ events, 94 $a_0(980)\pi$, 164 $\eta \sigma$ and
86 $a_2(1320)\pi$.
Log likelihood improves by 58 for 8 additional fitting parameters.
There is flexibility in the fit, and the total number of events can
be increased from 533 to 1182 if log likelihood is allowed to get
worse by 2 for every channel, corresponding to an overall 4 standard
deviation change.
This is a cautious limit taking account of the fact that the
components fitting this mass range are uncertain.
The corresponding branching fraction is
$BR(J/\Psi \to \gamma X).BR(X \to \pi ^+\pi ^- \eta) = 0.38 \times
10^{-4}$.
No statement can be made about its possible coupling to $K\bar K\pi$,
since the mass range of the present data extends only to 1800 MeV.

Huang and Zhu \cite {Zhu} point out that its mass, total width,
production rate and decay pattern favour its assignment as the
second radial excitation of $\eta '$.
More precisely, the small branching fraction in $\eta \pi \pi$
favours a mixing angle between $n\bar n$ and $s \bar s$ close to
that of the $\eta '$.
The $\eta'(958)$, $\eta (1440)$ and $X(1835)$ would then form a Regge
trajectory with a spacing in $M^2$ of $\sim 1.22$ GeV$^{2}$, close
to that of other well establshed trajectories \cite {Review}.
The peak observed by BES2 at 2240 MeV in $\phi \phi$ would fit
naturally as the fourth member of this sequence \cite {eta2240}.
However, resonant phase variation needs to  be established for
$\eta (1835)$ and the $\phi \phi$ peak.

Ma \cite {Ma} tries to interpret $B(1835)$ as a baryonium state
associated with the threshold peak observed in $\bar pp$ in
BES data \cite {ppthresh}.
This interpretation leads to a decay stronger to $\eta \pi \pi$ than
$\eta '\pi \pi$, and is clearly ruled out by the present data.

\section {Interpretation of $\eta (1440)$ and $f_1(1510)$}
If there is just one $\eta (1440)$ in the mass range 1400--1500 MeV,
it either has a large SU(3) singlet $q\bar q$ component or a glueball
component.
Either would explain its strong production in $J/\Psi$ radiative
decays, where $c\bar c$ appear in a singlet combination.
It could likewise be produced strongly in $\bar pp$ annihilation.
However the $K^*K$ channel is SU3 octet, so the $\eta (1440)$
(or $\eta (1475)$) cannot be purely singlet.

It is well known that isolated $s\bar s$ states such as
$f_2(1525)$ are produced weakly in $\bar pp$ annihilation.
If the $\eta (1475)$ is to be produced strongly  in $\bar pp$
annihilation, it should have a substantial $n\bar n$ component, but
that is inconsistent with decays purely to $K^*\bar K$.
The $\eta (1405)$ decays much more strongly to $\kappa K$ and
$f_0(980)\eta$ than to $\eta \sigma$ + $a_0(980)\pi$.
This would not arise naturally if it is a pure $n\bar n$ state.
So the conclusion is that a substantial $s\bar s$ (or gluball)
component must be present in $\eta (1440)$ or $\eta (1405)+\eta
(1475)$.

The existence of $\eta (1295)$ depends primarily on data in three
papers on $\pi ^-p \to (\eta \pi \pi)n$ and one on $\pi ^-p \to
(K\bar K\pi)n$ \cite {PDG}.
In this process, there should be a clear distinction between
$\eta (1295)$ which decays via the S-wave and $f_1(1285)$ which decays
via the P-wave.
If production is limited to a single exchange process, these are
immediately distinguishable.
In $\bar pp$ annihilation the situation is much more complex, since
there are four combinations of $\eta \pi \pi$ produced in every
event \cite {Nana}, making the separation between $\eta (1295)$
and $f_1(1285)$ much more difficult.
Those reviews which argue against the existence of $\eta (1295)$
base their arguments on the fact that there could be more than
one exchange mechanism in $\pi ^-p$ production data, leading to
possible confusion between the flat angular distribution from
$\eta (1295)$ and rank 2 processes involving $f_1(1285)$.
It is unfortunate that the papers of primary importance
published on $\eta (1295)$ do not give quantitative estimates of its
significance.
If the data still survive, it would be valuable if the
groups concerned would publish values of log likelihood or $\chi^2$
showing quantitatively how their fit changes when the $\eta (1295)$ is
omitted.
This information would sharpen the arguments for or against
$\eta (1295)$.

In the data presented here, most of the peak near 1285 MeV certainly
is due to $f_1(1285)$; a free fit including $\eta (1295)$
improves log likelihood only by 3.7 for four parameters fitted to
magnitudes and phases of decay amplitudes to $a_0(980)\pi$ and
$\eta \sigma$.
Seven events are fitted to $\eta (1295)$ compared with 230 to
$f_1(1285)$.
So the evidence for $\eta (1295)$ in present data is not significant.

Is the $\eta (1440)$ really split into two components?
Although fit A optimises $\eta (1405)$ at 1429 MeV, a persistent
feature of data fitted to $\eta (1405)$ in other analyses is a
mass 1405 MeV in $\eta \pi \pi$ and 1414 MeV in $\kappa \bar K$.
As shown in Section 5.3, data from $\bar pp$ annihilation require
corrections to these masses for the phase space in the production
process.
This shifts the mass fitted to $\eta \pi \pi$ data up by
4.7 MeV and that in $K\bar K\pi$ data by 9.7 MeV.
Weighting these with the intensities fitted to $\eta \pi \pi$ and
$\kappa \bar K$ in Table 2, the mean mass of $\eta (1405)$ becomes
1421 MeV.
There is then a discrepancy in mass of $18\pm 5$ MeV with the mass
in fit A with a single resonance.
This is not convincing evidence for two separate resonances near
1440 MeV.
However, it must be said that one cannot move the $\eta (1440)$ much
below 1439 MeV without a poor fit to the peak in $K\bar K\pi$.
How could the difference of 18 MeV be explained between
$\eta \pi \pi$ data and $K\bar K$?

Klempt \cite {zero}  has pointed out
a possible explanation.
A radial excitation has a node in its wave function causing a zero in
decays to $a_0\pi$ in this general mass range.
Firstly this attenuates the $\eta \pi \pi$ signal and may explain why
it is weaker than $\kappa \bar K$ and $K^*\bar K$.
Secondly, such a node introduces a further $s$-dependence into
$\eta \pi \pi$.
Suppose the amplitude for the $a_0\pi$ channel is multiplied by a
factor $(s_0 - s)/s_0$, where $s_0$ is the location of the zero in
the amplitude.
A resonance at 1439 MeV can have a peak at 1410 MeV if there is a zero
in the amplitude at 1.47 GeV, just where there is a dip in the $\eta
\pi \pi$ mass spectrum, Fig. 7.
However, the peak is attenuated more on its upper side than its lower
side and the peak is somewhat asymmetric.
Tests on the present data show that this explanation does cure the
10 MeV discrepancy between $\eta \pi \pi$ data and the fit near
1500 MeV.
However, a factor 10 higher statistics are needed to test the idea
precisely.
It cannot be tested in $\bar pp$ annihilation because of the
four different charge combinations.
Zeros would appear at different masses in $\eta \sigma$ and
$\kappa \bar K$, depending on matrix elements, leading to a
potentially complicated line-shape with peaks at different masses in
different channels.

There is one further possible complication.
The dominant decay of $\eta (1440)$ is through $K\bar K\pi$.
There is inevitably some rescattering of $K\bar K$ through
$a_0(980)$ to the $\eta \pi \pi$ channel via what are called
triangle graphs.
This was discussed in some detail in Ref. \cite {Nana}.
It will have the effect of increasing the $K\bar K\pi$ decay
branching fraction of $\eta (1440)$.

Let us turn to $\eta (1890)$.
The general pattern of the broad components in the $\eta \pi \pi$ mass
projection of Fig. 7 is that there are two components peaking at
1890 and 1600 MeV.
It is possible that the latter is resonant.
In the $\gamma 4\pi$ data there is evidence for a resonance at 1970
MeV with a width of 210 MeV.
If the parameters of $\eta (1890)$ are changed to those values, log
likelihood for $\eta \pi \pi$ data is worse by 42.
This is a modest amount, in view of the systematic
uncertainties about how to fit the $\eta \pi \pi$ mass range
above 1600 MeV and the strong $K^*\bar K^*$ threshold nearby.
It is therefore possible that these two signals arise from the same
resonance, though this cannot be proved at the moment.

Consider next $f_1(1420)$ and $f_1(1510)$.
In Fig. 8(a), one sees  that the effect of $f_1(1510)$ is to produce a
slight bump at 1490 MeV and extend the tail of $f_1(1420)$.
Longacre \cite {Ron} has suggested that $f_1(1420)$ may be an abnormal
state (or `molecule' in current terminology) with an $L=1$ pion
orbiting a $K\bar K$ S-wave combination.
In this picture, there are final state interactions from two
combinations of $K^*$ and a further final-state interaction between
the two S-wave kaons.
My suggestion is that these final state interactions enhance this
state strongly at the lowest $K\bar K$ masses, but the rapid rise of
the $K^*\bar K$ phase space with momentum produces a long tail
extending into the mass region of what is called $f_1(1510)$.

It must be said in favour of the data of Aston et al. that they
identified the $h_1(1370)$ $J^{PC} = 1^{-+}$ state and it was
subsequently confirmed by Crystal Barrel data \cite {h1370}.
Furthermore, there is a $2^{-+}$ peak in their data which they did
not claim as a resonance, but was subsequently identified as a
resonance by Crystal Barrel \cite {et21860} and confirmed by
WA102 \cite {WA102}.
The conclusion is that the data of Aston et al. appear reliable.
The primary question is one of interpretation of their $J^{PC} =
1^{++}$ signal.
They did not include $f_1(1420)$ in their fit.
So it is possible that the $f_1(1420)$ has a more extended line-shape
than has previously been fitted.

\section {Summary}
The main experimental results are the observation
of three peaks in the $\eta \pi \pi$ mass spectrum: a narrow peak
at 1415 MeV, a shoulder at 1550 MeV and a broad peak at $\sim
1850$ MeV.
In $K\bar K\pi$, there is an asymmetric peak at $\sim 1450$ MeV.
The asymmetry is due to the $s$-dependence of the $K^*\bar K$
channel;
its Dalitz plot immediately identifies a dominant $K^*\bar K$ decay.
The intensity for that decay increases rapidly from its threshold
at $\sim 1390$ MeV as $k^3$, where $k$ is the momentum of decay
$K$ and $K^*$ in the $K^*K$ rest frame.

An essential point from the present analysis is that one must
include this $s$-dependence correctly into the Breit-Wigner
amplitude.
This reduces the mass of $\eta (1475)$ to 1440 MeV.

When analysing $\bar pp$ data, it is necessary to allow for the
variation of phase space for production of $\eta \pi \pi$ and
$\kappa \bar K$ with mass.
This corrects the mass of $\eta (1405)$ upwards by 4.7 MeV in
$\eta \pi \pi$ and by 9.7 MeV in $\kappa \bar K$.
As a result, the mean mass difference between the possible
$\eta (1405)$ and $\eta 1440)$ is no larger than $18 \pm 5$ MeV.

There is a dominant broad $0^-$ amplitude in $\eta \pi \pi$.
When this is included into the partial wave fit, interference
with a single $\eta (1440)$ can account naturally for the observed
peaks at 1415 MeV in $\eta \pi \pi$ and at 1450 MeV in $K\bar K\pi$.
There is no evidence for two separate $\eta (1405)$ and $\eta (1475)$
from the present data.

The $\eta (1835)$ is not observed in the present $\eta \pi \pi$ data,
and an upper limit has been derived for the branching fraction for
its production in $J/\Psi $ decays.
This limit is a factor $5.7 \pm 1(stat) \pm 1(syst)$ smaller
than the $\eta (1835)$ observed in the $\eta '\pi \pi$ channel.
The natural explanation is that $\eta (1835)$ is a
radial excitation of $\eta '$ and $\eta (1440)$.

As an aid to further work, the fully annotated Fortran code for the
broad $0^-$ component and $\eta (1440)$ is available from the author.
This code includes numerical values of $g^2$ from the present work,
but they would need to be re-optimised if further data become
available.

\vspace{0.5cm}
I wish to acknowledge financial support from the Royal Society
and major help from members of the BES collaboration in processing
data and running the Monte Carlo simulation of acceptance and
backgrounds.


\begin{thebibliography}{99}
\bibitem {PDG} Particle Data Group (PDG), Physics Letters {\bf B667} 1
(2008)  
\bibitem {Fukui} S. Fukui {\it et al.}, Phys. Lett. B {\bf 267} 293
(1991) 
\bibitem {Alde} D. Alde {\it et al.} (GAMS Collaboration), Yad.
Fis. {\bf 60} 458 (1997), translated in Part. At. Nuclei {\bf 60} 386
(1997)  
\bibitem {Manak} J.J. Manak {\it et al.} (E852 Collaboration),
Phys. Rev. {\bf D62} 012003 (2000)  
\bibitem {Adams} G.S. Adams {\it et al.} (E852 Collaboration), Phys.
Lett. {\bf B516} 264 (2001)   
\bibitem{Nana} A.V. Anisovich {\it et al.}, Nucl. Phys. A {\bf 690}
567 (2001)   
\bibitem {KlemptZ} E. Klempt and A. Zaitsev, Phys. Rep. {\bf 454} 1
(2007)   
\bibitem {Masoni} A. Masoni, C. Cicalo and G.L. Usai, J. Phys. G: Nucl.
Phys. {\bf 32} R293 (2006)   
\bibitem {Morningstar} C.J. Morningstar and M. Peardon,
Phys. Rev. {\bf D60} 034509 (1999)    
\bibitem {Escribano} R. Escribano, arXiv: hep-ph/0802.3909 
\bibitem {X1835} M. Ablikim {\it et al.} (BES Collaboration), Phys.
Rev. Lett. {\bf 95} 262001 (2005)   
\bibitem {DetectA} J.Z. Bai {\it et al.} (BES Collaboration), Nucl.
Instr. Methods {\bf A344} 319 (1994) 
\bibitem {DetectB} J.Z. Bai {\it et al.} (BES Collaboration), Nucl.
Instr. Methods {\bf A458} 627 (2001) 
\bibitem {Edwards} D. Edwards {\it et al.} (Crystal Ball
Collaboration), Phys. Rev. Lett. {\bf 51} 859 (1983)  
\bibitem {g4pi} D.V. Bugg, accompanying paper arXiv:0907.3021  
\bibitem {BaiM3} Z. Bai {\it et al.} (Mark III Collaboration), Phys.
Rev. Lett. {\bf 65} 2507 (1990) 
\bibitem{wpp} M. Abklikim {\it et al.} (BES Collaboration), Phys. Lett.
B {\bf 598} 149 (2004)  
\bibitem {KappBES} M. Ablikim {\it et al.} (BES Collaboration), Phys.
Lett. {\bf B633} 681 (2006)   
\bibitem {JKappa} D.V. Bugg, Phys. Lett. {B632} {\bf 471} (2006)  
\bibitem {Aston} D. Aston {\it et al.} (LASS Collaboration),
Nucl. Phys. B {\bf 296} 493 (1988) 
\bibitem{ZouB} B.S. Zou and D.V. Bugg, Phys. Rev. {\bf D48} R3948
(1993)   
\bibitem{sigpole} D.V. Bugg, J. Phys. {\bf G34}  151 (2007) 
\bibitem {formulae} B.S. Zou and D.V. Bugg, Eur. Phys. J {\bf A 16} 537
(2003) 
\bibitem {a0980} D.V. Bugg, Phys. Rev. D {\bf 78} 074023 (2008)
\bibitem {BSZ} D.V. Bugg, A. V. Sarantsev and B.S. Zou, Nucl. Phys.
B {\bf 471} 59 (1996)  
\bibitem {phipp} M. Ablikim {\it et al.} (BES Collaboration) Phys.
Lett. {\bf B607} 243 (2005) 
\bibitem {fA1510} D. Aston {\it et al.} (LASS
Collaboration) Phys. Lett. {\bf B201} 573 (1988) 
\bibitem {Anisovich} A.V. Anisovich {\it et al.} Phys. Lett. {\bf B472}
168 (2000) 
\bibitem {Baillon} P. Baillon {\it et al.}, Nu. Cim. {\bf 50A} 383
(1967) 
\bibitem {Amsler} C. Amsler {\it et al.} (Crystal Barrel
Collaboration), Phys. Lett. {\bf B358} 389 (1995) 
\bibitem {CBAR} A.V. Anisovich {\it et al.}
(Crystal Barrel Collaboration), Phys. Lett. B {\bf 477} 19 (2000) 
\bibitem {Zhu} T. Huang and S-L. Zhu, Phys. Rev. {\bf D73} 014023
(2006)    
\bibitem {Review} D. V. Bugg, Phys. Rep. {\bf 297} 257 (2004) 
\bibitem {eta2240} M. Ablikim {\it et al.}, (BES Collaboration),
Phys. Lett. {\bf B662} 330 (2008)  
\bibitem {Ma} Y-L. Ma, arXiv: 0803.3764  
\bibitem {ppthresh} M. Ablikim {\it et al.} (BES
Collaboration), Phys. Rev. Lett. {\bf 95} 262001 (2005)   
\bibitem{zero} E. Klempt, hep-ph/0409148 and Beijing 2004, ICHEP 2004,
vol. 2, p1082  
\bibitem {Ron} R.S. Longacre, Phys. Rev. {\bf D42} 874 (1990)  
\bibitem {h1370} A. Abele {\it et al.} (Crystal Barrel
Collaboration), Phys. Lett. {\bf B415} 280 (1997)  
\bibitem{et21860} J. Adomeit {\it et al.} (Crystal Barrel
Collaboration), Z. Phys. C {\bf 71} 227 (1996)   
\bibitem {WA102} D. Barberis {\it et al.} (WA102 Collaboration),
Phys. Lett. {\bf B413} 217 (1997)  
\end{thebibliography}
\end {document}